\documentclass[seceq]{ptptex}
\usepackage{amsmath,amsfonts,amssymb,wrapfig}
\usepackage[dvips]{graphicx}
\def\vecx{\mbox {\boldmath $x$}}
\def\vecy{\mbox {\boldmath $y$}}
\def\veck{\mbox {\boldmath $k$}}
\def\vecq{\mbox {\boldmath $q$}}
\def\ha{{1\over 2}}
\def\lan{\langle}
\def\ran{\rangle}
\def\bra#1{\lan#1|}
\def\ket#1{|#1\ran}

\title{Perturbative Formulation of Pure Space-Like Axial Gauge QED
with Infrared Divergences Regularized by Residual Gauge Fields }

\author{ Yuji {\sc Nakawaki} and Gary {\sc McCartor}$^*$
}

\inst{Division of Physics and Mathematics, Faculty of Engineering \\
 Setsunan University, Osaka 572-8508, Japan
\\
$^*$Department of Physics, Southern Methodist University  \\
Dallas TX 75275,  USA}

\recdate{
}

\abst{We construct a new perturbative formulation of pure space-like axial
gauge QED in which the inherent infrared divergences are regularized by
residual gauge fields.  For that purpose we perform our calculations in 
coordinates $x^{\mu}=(x^+,x^-,x^1,x^2)$, where $x^+=x^0\sin{\theta}+x^3\cos
{\theta}$ and $x^-=x^0\cos{\theta}-x^3\sin{\theta}$.  $A_-=A^0\cos{\theta}+A^3
\sin{\theta}=n{\cdot}A=0$ is taken as the gauge fixing condition. That 
framework allows us to show that the residual gauge fields are not canonical 
variables in a pure 
space-like axial gauge formulation, although they are necessary ingredients of 
the system. To overcome that difficulty we first construct, in the region 
$0{\leqq}\theta<\frac{\pi}{4}$, a temporal gauge formulation in which $x^-$ is 
taken as the evolution parameter and $A_+,A_i(i=1,2)$ and their canonical 
conjugate momenta are independent canonical variables so that we can quantize 
them by canonical methods. As a result we can obtain the commutation relations 
of the $x^-$-independent Nakanishi-Lautrup field $B$, which we extrapolate 
into the 
axial region $\frac{\pi}{4}<\theta<\frac{\pi}{2}$, where $x^+$ is taken as the 
evolution parameter. We obtain a formal solution of the gauge field 
equations with the relevant integration constants specified on the basis of 
the free gauge fields in the temporal formulation. Then we make use 
of the solution to specify the integration constants appearing when we solve 
the axial gauge constraint equations. It turns out that, because the residual 
gauge fields are multiplied by the inverse Laplace operator 
$\frac{1}{n_-\partial_+^2+\partial_{\bot}^2}$, which becomes singular in the 
axial region, they give rise to vanishing contributions to the equal 
$x^+$-time canonical commutation relations. For this reason, we can uniquely 
quantize the physical degrees of freedom and we do not obtain any extra terms 
in the equation for the Fermi field in spite of having introduced the residual 
gauge fields.  We have also obtained the conserved Hamiltonian by 
integrating the canonical energy-momentum tensor $T_{\mu \nu}$ around a 
suitable contour. The Hamiltonian consists 
of the usual terms, with the residual gauge fields in $A_\mu$ as regulators, 
plus the Hamiltonian for the residual gauge fields which is determined by 
integrating a density over the hyperplane $x^-$ = constant.  We show in detail 
that, in perturbation theory, infrared divergences resulting
 from the residual gauge fields cancel infrared divergences resulting
 from the physical parts of the gauge field. As a result we obtain the 
gauge field propagator prescribed by Mandelstam and Leibbrandt. By taking the 
limit $\theta {\to} \frac{\pi}{4}$ we can construct the light-cone 
formulation which is free from infrared difficulty.  With that analysis 
complete, we perform a successful calculation of the one loop electron self 
energy, something not previously done in light-cone quantization and 
light-cone gauge.}   

\begin{document}

\maketitle

\section{Introduction}
Axial gauges, $n^{\mu}A_{\mu}=0$, specified by a constant vector $n^{\mu}$, 
have been used recently in spite of their lack of manifest Lorentz covariance.
\cite{rf:1} One reason is that the Faddeev-Popov ghosts decouple 
in the axial gauge (AG) formulations.\cite{rf:2} For that reason we will not 
need to discuss the Faddeev-Popov ghosts in this paper and the term ghost will 
always refer to residual gauge (RG) fields which are introduced as integration 
constants in solutions to constraint equations. The case $n^2=0$, the 
light-cone gauge, has been extensively used in the light-cone field theory in 
attempts to find nonperturbative solutions to QCD.\cite{rf:3} The same 
formulation is often used in phenomenological calculations where the low 
energy properties of the theory are parameterized in light-cone wave 
functions.  Traditionally, constraints in AG formulations are solved to 
eliminate dependent fields in terms of physical fields. That elimination 
requires one to utilize inverse 
derivatives which introduce so-called spurious singularities. It was first 
pointed out by Nakanishi\cite{rf:4} that there exists an intrinsic difficulty 
in the AG formulations, the solution to which requires an indefinite metric; 
even in QED. It was also noticed that in order to bring perturbative 
calculations done in the light-cone gauge into agreement with calculations 
done in covariant gauges, spurious singularities of the free gauge field 
propagator have to be regularized, not as principal values, but according 
to the Mandelstam-Leibbrandt (ML) prescription\cite{rf:5} in such a way that 
causality is preserved. Shortly afterwards, Bassetto et al.\cite{rf:6} found 
that the ML form of the propagator is obtained in the light-cone gauge 
canonical formalism in ordinary coordinates if one introduces a Lagrange 
multiplier field and its conjugate as RG degrees of freedom. 

It is now established\cite{rf:7} that AG formulations are not ghost free, 
contrary to what was 
originally expected, and is still sometimes claimed. Nevertheless, the problem 
of finding the correct method by which to introduce 
RG fields into pure space-like AG formulations has not been completely solved. 
A first step toward solving this problem was made by McCartor 
and Robertson in the light-cone formulation of QED.\cite{rf:8} In previous 
papers,\cite{rf:9} we considered 
the solvable Schwinger model and learned ways to introduce 
and to quantize RG fields. In this paper we apply these methods to the more 
realistic problem of QED 
and construct an extended Hamiltonian formalism of QED in the 
$\pm$-coordinates $x^{\mu}=(x^+,x^-,x^1,x^2)$, where 
$x^+=x^0\sin{\theta}+x^3\cos{\theta},\; 
x^-=x^0\cos{\theta}-x^3\sin{\theta}$
and the space-like constant vector $n$ is specified to be 
$n^{\mu}=(n^+,n^-,n^1,n^2)=(0,1,0,0)$ 
so that the gauge fixing condition is pure space-like. The same 
framework was used by others to analyze 
two-dimensional models.\cite{rf:10} 
We show that if we properly choose constituent 
fields, then we can construct the extended Hamiltonian formalism of QED in 
quite the same manner as the Schwinger model.

In spite of the common use of light-cone gauge (with $A_+$ treated as a 
constrained field) and light-cone quantization, the one loop electron self 
energy has never been successfully calculated in that formulation. To our 
knowledge, that most fundamental of loop calculations has been successfully 
calculated in three formulations of light-cone quantization:  Morara and 
Soldati\cite{rf:7} calculated it in the gauge $A_+ = 0$;  Langnau and 
Burkardt\cite{rf:10a} calculated it in Feynman gauge but did not start the 
calculation on the light-cone; in \cite{rf:10b} it was calculated in Feynman 
gauge starting from the light-cone and also calculated in light-cone gauge, but
 with a higher derivative regulator so that $A_+$ was a degree of freedom, not 
a constraint.  But it has not been calculated in light-cone gauge with $A_+$ 
treated as a constraint.  The problems that prevent such a calculation can be 
glimpsed in \cite{rf:10a} and are discussed in detail in \cite{rf:10b}.  
Basically, the infrared singularities are too strong to allow a successful 
calculation.  In the present paper we shall show that with the RG fields in 
place, the infrared singularities are softened in such a way as to allow a 
successful calculation.

The paper is organized as follows. In ${\S}2$, we construct the temporal gauge
 (TG) formulation and obtain the commutation relations of the Nakanishi-Lautrup
 field $B$, which are then extrapolated into the axial region. In ${\S}3$ we 
consider the dipole ghost field and the constituent fields corresponding to 
those in the $n{\cdot}A=0$ gauge Schwinger model. We then express $A_{\mu}$ in 
terms of these constituent fields. In ${\S}~4$ we obtain the translational 
generators by applying McCartor's method.\cite{rf:11} In ${\S}~5$ perturbation 
theory is developed and we show that if we define the singularities resulting 
from inversion of a hyperbolic Laplace operator as principal values, the worst 
infrared divergences resulting from physical parts 
of the gauge fields are cancelled by infrared divergences from the RG parts.  
We then find that the remaining infrared divergences are regularized as the ML 
form of gauge field propagator. In that section we also perform a calculation 
of the one loop electron self energy.  ${\S}~6$ is devoted to concluding 
remarks.

We use the following conventions: \\
Greek indices ${\mu},{\nu},{\cdots}$ will take the values 
$+,-,1,2$ and label the component of a given four-vector (or tensor) in 
the $\pm$ coordinates; \\
Latin indices $i,j,{\cdots}$ will take the values 
$1,2$ and label the $1,2$ component of a given four-vector (or tensor) in 
the $\pm$ coordinates; \\
the Einstein convension of sum over repeated indices will be always used; 
\begin{eqnarray*}
&\vecx^{\pm}=(x^{\pm},x^1,x^2),\;\vecx_{\bot}=(x^1,x^2),\;
d^2\vecx_{\bot}=dx^1dx^2,\;d^3\vecx^{\pm}=dx^1dx^2dx^{\pm}& \\
& \veck_{\pm}=(k_{\pm},k_1,k_2),\;d^3\veck_{\pm}=dk_1dk_2dk_{\pm}& \\
&g_{--}=\cos2{\theta}, \quad g_{-+}=g_{+-}=\sin2{\theta},
 \quad g_{++}=-\cos2{\theta}&  \\
&g_{-i}=g_{i-}=g_{+i}=g_{i+}=0, \quad g_{ij}=-{\delta}_{ij}. & \\
&{\gamma}^+={\gamma}^0\sin {\theta}+{\gamma}^3\cos {\theta},\quad
{\gamma}^-={\gamma}^0\cos {\theta}-{\gamma}^3\sin {\theta}.&
\end{eqnarray*}

\section{Temporal gauge formulation of $n{\cdot}A=0$ gauge QED}

In this section we keep $\theta$ in the temporal region $0{\leqq}\theta<
\frac{\pi}{4}$ and choose $x^-$ as the evolution parameter. This enables us to 
show that RG fields are necessary ingredients in the canonical quantization of 
$n{\cdot}A=0$ gauge QED. 

The $n{\cdot}A=0$ gauge QED is defined by the Lagrangian
\begin{equation}
L=-\frac{1}{4}F_{{\mu}{\nu}}F^{{\mu}{\nu}}-B(n{\cdot}A) 
+\bar{\Psi}(i{\gamma}^{\mu}D_{\mu}-m){\Psi}  
\label{eq:2.1}
\end{equation} 
where $D_{\mu}={\partial}_{\mu}+ieA_{\mu}$ and  $B$ is the Lagrange 
multiplier field, that is, the Nakanishi-Lautrup field in noncovariant 
formulations.\cite{rf:12} From (\ref{eq:2.1}) we derive the field equations
\begin{equation}
{\partial}_{\mu}F^{{\mu}{\nu}}=n^{\nu}B +J^{\nu}, \quad 
J^{\nu}=e\bar{\Psi}{\gamma}^{\nu}{\Psi} \label{eq:2.2}
\end{equation}
\begin{equation}
(i{\gamma}^{\mu}D_{\mu}-m){\Psi}=0, \label{eq:2.3}
\end{equation}
and the gauge fixing condition  $n{\cdot}A=0$. The field equation 
of $B$,
\begin{equation}
(n{\cdot}{\partial})B={\partial}_-B=0, \label{eq:2.4}
\end{equation}
is obtained by operating on (\ref{eq:2.2}) with ${\partial}_{\nu}$.

Canonical conjugate momenta are defined to be
\begin{eqnarray}
&{\pi}^+=\frac{{\delta}L}{{\delta}{\partial}_-A_+}=F_{-+}, \quad
{\pi}^-=\frac{{\delta}L}{{\delta}{\partial}_-A_-}=0, \quad
{\pi}^i=\frac{{\delta}L}{{\delta}{\partial}_-A_i}=F^-_{\;\;i},& \nonumber \\
&{\pi}_{\scriptscriptstyle B}=\frac{{\delta}L}{{\delta}{\partial}_-B}=0, \quad
{\pi}_{\scriptscriptstyle \Psi}=\frac{{\delta}L}{{\delta}{\partial}_-{\Psi}}
=i\bar{\Psi}{\gamma}^-  \label{eq:2.5}
\end{eqnarray}
and the Gau\ss \ law constraint is described in terms of them as
\begin{equation}
{\partial}_+\pi^+ +{\partial}_i\pi^i=B+J^-. \label{eq:2.6}
\end{equation}
We see from (\ref{eq:2.5}) and (\ref{eq:2.6}) that in the TG formulation we 
have three pairs of bosonic canonical variables, $A_+,\pi^+,\;
A_i,\pi^i\;(i=1,2)$, which indicates indispensability of degrees of freedom 
other than the physical ones in constructing the canonical formulation. 
By using the canonical equal $x^-$-time quantization conditions imposed on the 
independent canonical variables and the expression of $B$ obtained from 
(\ref{eq:2.6}) we easily obtain the commutation relations of $B$:
\begin{eqnarray}
&[B(x),A_{\mu}(y)]=i{\partial}_{\mu}\delta^{(3)}(\vecx^+-\vecy^+), \quad 
[B(x),\pi^+(y)]=[B(x),\pi^i(y)]=0,& \nonumber \\
&[B(x),B(y)]=0, \quad [B(x),\Psi(y)]=e\delta^{(3)}(\vecx^+-\vecy^+)\Psi(y). &
\label{eq:2.7}
\end{eqnarray}
It is important to note here that, due to (\ref{eq:2.4}), these commutation 
relations hold even when $x^-{\ne}y^-$.

The Hamiltonian, that is, the translational generator for the $x^-$-direction, is given by 
\begin{eqnarray}
P_-=\int\hspace*{-1mm}d^3\vecx^+T_-^{\;-}&=&\int\hspace*{-1mm}d^3\vecx^+
\{2^{-1}((\pi^+)^2+n_-^{\;-1}
(\pi^i-n_+F_{+i})^2+n_-(F_{+i})^2+(F_{12})^2) \nonumber \\
&+&\bar{\Psi}(m-i\gamma^+\partial_+-\gamma^i\partial_i)
\Psi+J^{\mu}A_{\mu} \}
\end{eqnarray}
so that it is straightforward to develop an $x^-$-time ordered perturbation 
theory. It was shown previously that the free gauge field is described as
\cite{rf:13} 
\begin{equation}
A_{\mu}=T_{\mu} -\frac{n_{\mu}}{{\nabla}_T^{\;\;2}}B+\partial_{\mu}\Lambda
\end{equation}
where $T_{\mu}$ is a free field satisfying
\begin{eqnarray}
&\square T_{\mu}=0, \quad \partial^{\mu}T_{\mu}=0, \quad n{\cdot}T=0,
\label{eq:2.10}& \\
&[T_{\mu}(x),T_{\nu}(y)]=i(-g_{\mu\nu}+\frac{n_{\mu}\partial_{\nu}
+n_{\nu}\partial_{\mu}}{\partial_-}-n^2\frac{\partial_{\mu}\partial_{\nu}}
{\partial_-^{\;2}})D(x-y) \label{eq:2.11} &
\end{eqnarray}
and
\begin{eqnarray}
&{\nabla}_T^{\;\;2}=\partial_1^{\;2}+\partial_2^{\;2}+n_-\partial_+^{\;2},&\\
&\Lambda=-\frac{1}{{\nabla}_T^{\;\;2}}
\left(C-n_-x^-B-\frac{n_-n_+}{{\nabla}_T^{\;\;2}}\partial_+B\right). 
\label{eq:2.13}&
\end{eqnarray}
Note in (\ref{eq:2.11}) that $D(x-y)$ is the commutation function of the free 
massless field and  that the inverse derivative $\partial_-^{\;-1}$ is 
defined by
\begin{equation}
\partial_-^{\;-1}f(x)=\frac{1}{2}\int_{-\infty}^{\infty}dy^-\varepsilon
(x^--y^-)f(x^+,y^-,\vecx_{\bot}) 
\end{equation} 
which imposes, in effect, the principal value regularization. 
Note furthermore that $C$ in (\ref{eq:2.13}) is the conjugate of $B$ and 
satisfies
\begin{equation}
\partial_-C=0, \quad [C(x),C(y)]=0, \quad [B(x),C(y)]=-i{\nabla}_T^{\;\;2}
\delta^{(3)}(\vecx^+-\vecy^+).
\end{equation}

It is useful to point out here that the physical degrees of freedom are carried by 
$T_{\mu}$, as is seen from (\ref{eq:2.10}), and the remaining degrees of 
freedom are carried by the pair of functions $B$ and $C$. It should be noted in particular that solving the on-mass-shell condition
\begin{equation}
0=p^2=n_-p_-^{\;2}+2n_+n_-p_+p_--n_-p_+^{\;2}-p_{\bot}^{\;2}, 
\quad (p_{\bot}^{\;2}=p_1^{\;2}+p_2^{\;2})
\end{equation}
yields
\begin{equation}
p_-=\frac{\sqrt{p_+^{\;2}+n_-p_{\bot}^{\;2}}-n_+p_+}{n_-}, 
\quad p^-=\sqrt{p_+^{\;2}+n_-p_{\bot}^{\;2}}, \quad
p^+=\frac{n_+\sqrt{p_+^{\;2}+n_-p_{\bot}^{\;2}}-p_+}{n_-}
\end{equation}
from which follows
\begin{equation}
\frac{1}{p_-}=\frac{n_+p_++\sqrt{p_+^{\;2}+n_-p_{\bot}^{\;2}}}
{p_{\bot}^{\;2}+n_-p_+^{\;2}}.
\end{equation}
Therefore we obtain the factors $(p_{\bot}^{\;2}+n_-p_+^{\;2})^{-1}$ from 
the physical operators. It turns out that they are 
canceled  by those resulting from $\frac{1}{{\nabla}_T^{\;\;2}}$, which 
is appled to $B$ and $C$ so that we obtain the following ML form of the 
$x^-$-ordered gauge field propagator 
\begin{equation}
\langle 0|T(A_{\mu}(x)A_{\nu}(y))|0\rangle =\frac{1}{(2\pi)^4}
\int\hspace*{-1mm}d^4kD_{\mu\nu}(k){\rm e}^{-ik{\cdot}(x-y)}
\end{equation}
where
\begin{equation}
D_{\mu\nu}(k)=\frac{i}{k^2+i\varepsilon}\left\{-g_{\mu\nu}+\frac{n_{\mu}k_{\nu}
+n_{\nu}k_{\mu}}{k_-+i\varepsilon{\rm sgn}(k_+)}-n^2\frac{k_{\mu}k_{\nu}}
{(k_-+i\varepsilon{\rm sgn}(k_+))^2}\right\}. \label{eq:2.20}
\end{equation}
We see from (\ref{eq:2.20}) that RG fields are indispensable to regularize 
singularities associated with the gauge fixing in such a way that causality is 
preserved in complex $k_-$ coordinates. Because the factors $(p_{\bot}^{\;2}
+n_-p_+^{\;2})^{-1}$, which becomes singular in the axial region due to 
$n_-<0$, drop out completely from the gauge field propagator, we can formally 
extrapolate the gauge field propagator into the axial region. It is expected 
from this that we can develop a perturbation theory free from infrared 
divergences in the AG formulation. In next section we consider introducing 
RG fields so as to regularize the infrared divergences in the AG formulation.

\section{Constituent operators in the axial gauge formulation}

\subsection{ general solution of the gauge field equation (\ref{eq:2.2})} 
We begin by obtaining the general solution of (\ref{eq:2.2}) in a way which is 
independent of the evolution parameter. Let $a_{\mu}(x)$ be a field satisfying 
\begin{equation}
\square a_{\mu}=J_{\mu}. \label{eq:3.1}
\end{equation}
Then it holds that 
\begin{equation}
A_{\mu}-\frac{\partial_{\mu}}{{\nabla}_{\bot}^{\;\;2}}\partial_iA_i
+\frac{n_{\mu}}{{\nabla}_T^{\;\;2}}B= 
a_{\mu}-\frac{\partial_{\mu}}{{\nabla}_{\bot}^{\;\;2}}\partial_ia_i, 
\quad ({\nabla}_{\bot}^{\;\;2}=\partial_1^{\;2}+\partial_2^{\;2})
\label{eq:3.2}
\end{equation}
which is verified as follows: If we apply D'Alembert's operator, which 
is described in our formulation as 
$$\square=n_-\partial_-^{\;2}+2n_+\partial_+\partial_--n_-\partial_+^{\;2}
-\nabla_{\bot}^{\;\;2},
$$
to both sides of (\ref{eq:3.2}), then we obtain the same result, 
$J_{\mu}-\frac{\partial_{\mu}}{{\nabla}_{\bot}^{\;\;2}}\partial_iJ_i$.
Therefore the difference between the left hand side and the right hand side 
is a solution of the free massless D'Alembert's equation. The solutions 
which tend to zero at spatial infinity ($|\vecx_{\bot}|{\to}\infty$) vanish 
identically .

In addition we can show that $a_{\mu}$ satisfies
\begin{equation}
\partial^{\mu} a_{\mu}=0. \label{eq:3.3}
\end{equation}
In fact operating on (\ref{eq:3.2}) with $\partial^{\mu}$ yields
\begin{eqnarray*}
&\partial^{\mu} A_{\mu}-\frac{\partial^{\mu}\partial_{\mu}}{{\nabla}_{\bot}
^{\;\;2}}\partial_iA_i
=\partial^{\mu} a_{\mu}-\frac{1}{{\nabla}_{\bot}^{\;\;2}}\partial_iJ_i& \\
&\therefore \quad \partial_i(\square A_i -\partial_i\partial^{\mu} A_{\mu} 
-J_i)=\partial_i(\partial^{\mu}F_{\mu i}-J_i)
=-{\nabla}_{\bot}^{\;\;2}\partial^{\mu} a_{\mu}. &
\end{eqnarray*}
The left hand side of the second line vanishes due to the spatial component 
of the gauge field equation and so does the right hand side.

>From (\ref{eq:3.2}) $A_{\mu}$ is described as
\begin{equation}
A_{\mu}=a_{\mu}-\frac{\partial_{\mu}}{{\nabla}_{\bot}^{\;\;2}}\partial_ia_i
+\frac{\partial_{\mu}}{\sqrt{-\nabla_{\bot}^{\;\;2}}}X -
\frac{n_{\mu}}{{\nabla}_T^{\;\;2}}B \label{eq:3.4}
\end{equation}
where we have denoted  $-\frac{1}{\sqrt{-\nabla_{\bot}^{\;\;2}}}
{\partial}_iA_i$ as $X$.  
Throughout this 
paper the operator $\sqrt{-\nabla_{\bot}^{\;\;2}}$ is understood as  $|\veck
_{\bot}|$ in the Fourier transforms of the relevant operators.
The minus component of (\ref{eq:3.4}) has to vanish identically due to the 
gauge fixing condition. Imposing $A_-=0$ gives rise to the following 
constraint 
\begin{equation}
\frac{\partial_-}{\sqrt{-\nabla_{\bot}^{\;\;2}}}X
=-\{a_--\frac{\partial_-}{{\nabla}_{\bot}^{\;\;2}}\partial_ia_i\}
+\frac{n_-}{{\nabla}_T^{\;\;2}}B \label{eq:3.5} 
\end{equation}
the first term of which is rewritten, due to $\partial_ia_i=\partial^+a_+ 
+\partial^-a_-$, as
\begin{eqnarray}
a_--\frac{\partial_-}{{\nabla}_{\bot}^{\;\;2}}\partial_ia_i
&=&-\frac{1}{\nabla_{\bot}^{\;\;2}}\{\square a_+ -\partial^+(\partial_+a_- -
\partial_-a_+)\} \nonumber \\
&=&-\frac{1}{\nabla_{\bot}^{\;\;2}}\{J_- -
\partial^+(\partial_+a_- -\partial_-a_+)\}. \label{eq:3.6}
\end{eqnarray}
Thus integrating (\ref{eq:3.5}) with respect to $x^-$ yields $X$ 
expressed as
\begin{equation}
\frac{1}{\sqrt{-\nabla_{\bot}^{\;\;2}}}X
= -\frac{\partial_-^{\;-1}}{\nabla_{\bot}^{\;\;2}}\{
\partial^+(\partial_+a_- -\partial_-a_+) -J_- \} +\Lambda \label{eq:3.7}
\end{equation}
where we have introduced 
integration constants involving $C$ in the same manner as in (\ref{eq:2.13}).  
Substituting (\ref{eq:3.7}) into $X$ in (\ref{eq:3.4}), we have 
$A_{\mu}$ expressed as
\begin{equation}
A_{\mu}=a_{\mu}-\frac{\partial_{\mu}}{\partial_-}a_- 
-\frac{n_{\mu}}{{\nabla}_T^{\;\;2}}B+\partial_{\mu}\Lambda. 
 \label{eq:3.8}
\end{equation}

It is straightforward to show that, because $\partial_-B=\partial_-C=0$, $X$ 
and $A_{\mu}$ satisfy the following equations: 
\begin{eqnarray}
&\frac{\partial^{\mu}\partial_{\mu}}{\sqrt{-\nabla_{\bot}^{\;\;2}}}\partial_-X
=-n_-B+\frac{\partial_-}{{\nabla}_{\bot}^{\;\;2}}\partial_iJ_i -J_-, 
\label{eq:3.9}& \\
&\partial^{\mu}A_{\mu}=C-n_-x^-B+\frac{n_-n_+}{{\nabla}_T^{\;\;2}}\partial_+B
 -(\partial_-)^{-1}J_-.& \label{eq:3.10}
\end{eqnarray}
We see from (\ref{eq:3.9}) that $X$ is nothing but the dipole ghost field.

\subsection{Commutation relations of the constituent operators }

In the axial region, where $\frac{\pi}{4}<\theta<\frac{\pi}{2}$ and $x^+$ is 
chosen as the evolution parameter, we obtain from (\ref{eq:2.1}) the canonical 
conjugate momenta
\begin{eqnarray}
&{\pi}^+=\frac{{\delta}L}{{\delta}{\partial}_+A_+}=0, \quad
{\pi}^-=\frac{{\delta}L}{{\delta}{\partial}_+A_-}=F_{+-}, \quad
{\pi}^i=\frac{{\delta}L}{{\delta}{\partial}_+A_i}=F^+_{\;\;i},& \nonumber \\
&{\pi}_{\scriptscriptstyle B}=\frac{{\delta}L}{{\delta}{\partial}_+B}=0, \quad
{\pi}_{\scriptscriptstyle \Psi}=\frac{{\delta}L}{{\delta}{\partial}_+{\Psi}}
=i\bar{\Psi}{\gamma}^+.  
\end{eqnarray}
If we impose the gauge fixing condition, $A_-=0$, on the equation for $\pi^-$, then we obtain a 
constraint
\begin{equation}
{\pi}^-=-{\partial}_-A_+ \label{eq:3.12}
\end{equation} 
in addition to the Gau\ss \ law constraint  
\begin{equation}
{\partial}_-{\pi}^-+{\partial}_i{\pi}^i=J^+. \label{eq:3.13}
\end{equation}
As a consequence, $A_+$ becomes a dependent variable so that there remain only 
$A_i,\pi^i\;(i=1,2)$ as independent bosonic canonical variables. Therefore, as 
quantization conditions we have only 
equal $x^+$-time commutation and/or anticommutation relations on the 
independent canonical variables; the nonvanishing ones are
\begin{equation}
[A_i(x),\pi^j(y)]=i\delta_{ij}\delta^{(3)}(\vecx^--\vecy^-), \quad 
\{\Psi(x),\bar{\Psi}(y)\gamma^+ \}=\delta^{(3)}(\vecx^--\vecy^-).
\end{equation}

The degrees of freedom of the system should not change when we move from the 
TG formulation to the AG one, because the field equations and the 
gauge fixing condition are the same. How can we resolve this paradox? The only 
answer is that $B$ and $C$ take the place of the canonical variables 
$A_+,\pi^+$ in the TG formulation but happen 
not to be canonical variables in the AG formulation. This reflects 
the fact that  $B$ and $C$ are introduced as $x^-$ independent static fields 
so that $x^+$ can not be the evolution parameter for them. 
Therefore, we cannot obtain their quantization conditions from the Dirac 
procedure.\cite{rf:14}. To supplement the insufficient quantization conditions, we 
assume that the commutation relations of $B$ given in (\ref{eq:2.7}) can be 
extrapolated into the axial region. This is a reasonable assumption, because 
we have introduced degrees of freedom for c-number residual gauge 
transformations and because $B$ generates them. However it is not 
straightforward to make the extrapolation because the Laplace operator ${\nabla}_T^{\;\;2}$ 
becomes hyperbolic so that its inverse gives rise to 
singularities. Note that $n_-={\rm cos}2{\theta}<0$ in the axial region. We 
regularize these singularities by the principal value method. Consequently we obtain vanishing 
contributions at equal $x^+$-time due to 
$({\nabla}_T^{\;\;2})^{-1}$. For example, we obtain
\begin{eqnarray}
&{}&[\frac{1}{{\nabla}_T^{\;\;2}}B(x),
\frac{1}{\sqrt{-{\nabla}_{\bot}^{\;\;2}}}X(y)]_
{x^+=y^+}=\frac{-i}{{\nabla}_T^{\;\;2}}
\delta^{(3)}(\vecx^+-\vecy^+)|_{x^+=y^+} \nonumber \\ 
&=&\frac{i}{(2\pi)^3}\int d^3\veck_+\frac{1}
{|\veck_{\bot}|^2+n_-k_+^{\;2}}
{\rm e}^{i\veck_{\bot}{\cdot}(\vecx_{\bot}-\vecy_{\bot})}=0, \label{eq:3.15} \\
&{}&[\frac{1}{{\nabla}_T^{\;\;2}}B(x),\Psi(y)]_
{x^+=y^+}=\frac{e}{{\nabla}_T^{\;\;2}}
\delta^{(3)}(\vecx^+-\vecy^+)\Psi(y)|_{x^+=y^+}  \nonumber \\
&=&\frac{-e}{(2\pi)^3}\int d^3\veck_+\frac{1}{|\veck_{\bot}|^2+n_-k_+^{\;2}}
{\rm e}^{i\veck_{\bot}{\cdot}(\vecx_{\bot}-\vecy_{\bot})}\Psi(y)=0 
\label{eq:3.16}
\end{eqnarray}
where we have made use of the fact that 
\begin{equation}
\int_{-\infty}^{\infty}dk_+\frac{1}{|\veck_{\bot}|^2+n_-k_+^{\;2}}=0 
\end{equation}
as a result of regularizing the singular integral by the principal value method.
In this way $B$ makes vanishing contributions to equal $x^+$-time commutation 
relations. 

To express $A_+$ in terms of the independent canonical variables, 
we integrate (\ref{eq:3.12}) and (\ref{eq:3.13}) with respect to $x^-$. By 
integrating (\ref{eq:3.13}) we obtain
\begin{equation}
{\pi}^-=-\partial_-A_+=\frac{1}{\partial_-}(J^+-{\partial}_i{\pi}^i) + 
{\rm constant}. \label{eq:3.18}
\end{equation} 
The integration constant is determined by comparing (\ref{eq:3.18}) with that 
given by $A_+$ in (\ref{eq:3.8}) and we get 
\begin{eqnarray}
&\partial_+a_- -\partial_-a_+=\frac{1}{\partial_-}(J^+-{\partial}_i{\pi}^i),
\label{eq:3.19}&  \\
&{\rm constant}=-\frac{n_-}{{\nabla}_T^{\;\;2}}
\partial_+B.& 
\end{eqnarray} 
Integrating (\ref{eq:3.18}) with respect to $x^-$ and then comparing the result
 with $A_+$ in (\ref{eq:3.8}), we obtain
\begin{equation}
A_+=-\frac{1}{\partial_-^{\;2}}(J^+-{\partial}_i{\pi}^i)
-\frac{n_+}{{\nabla}_T^{\;\;2}}B +\partial_+\Lambda. \label{eq:3.21}
\end{equation} 

We see from (\ref{eq:3.21}) that $A_+$ consists of the conventional physical 
operator plus RG operators. Furthermore, we see from (\ref{eq:3.7}) and 
(\ref{eq:3.19}) 
that $\partial_+a_- -\partial_-a_+$ plays fundamental roles in the axial gauge 
formulation. If we define $\tilde{\Sigma}$ by
\begin{equation}
\partial_+a_- -\partial_-a_+=\frac{1}{\partial_-}(J^+-\partial_i\pi^i)
=\sqrt{-{\nabla}_{\bot}^{\;\;2}}\tilde{\Sigma},
\end{equation} 
then $\partial^+\tilde{\Sigma}$ is expressed  as
\begin{equation}
\partial^+\tilde{\Sigma}
=-\frac{1}{\sqrt{-{\nabla}_{\bot}^{\;\;2}}}
\partial_-\partial_iA_i-\frac{n_-\sqrt{-{\nabla}_{\bot}^{\;\;2}}}
{{\nabla}_T^{\;\;2}}B +
\frac{1}{\sqrt{-{\nabla}_{\bot}^{\;\;2}}}J_-, \label{eq:3.23} 
\end{equation}
so that at $x^+=y^+$ we obtain 
\begin{equation}
[\tilde{\Sigma}(x),\partial^+\tilde{\Sigma}(y)]
=i\delta^{(3)}(\vecx^--\vecy^-),
\quad [\tilde{\Sigma}(x),\tilde{\Sigma}(y)]=[\partial^+\tilde{\Sigma}(x),
\partial^+\tilde{\Sigma}(y)]=0. \label{eq:3.24}
\end{equation}
Here we have assumed the equal $x^+$-time current commutation relations: 
\begin{equation}
[J^+(x),J^+(y)]=[J^+,J_-(y)]=[J_-(x),J_-(y)]=0. \label{eq:3.25}
\end{equation} 
In addition, both $\tilde{\Sigma}$ and $\partial^+\tilde{\Sigma}$ are gauge 
invariant so that due to (\ref{eq:2.7}) their four dimensional commutator with 
$B$ vanishes:
\begin{equation}
[\tilde{\Sigma}(x),B(y)]=0,\quad [\partial^+\tilde{\Sigma}(x),B(y)]=0. 
\label{eq:3.26}
\end{equation} 
Note in particular that, when we calculate the commutation relations of 
$\partial^+\tilde{\Sigma}$ with $X$ and $\Psi$, we do not obtain any 
$\delta^{(3)}(\vecx^+-\vecy^+)$ contributions due to (\ref{eq:3.15}) and 
(\ref{eq:3.16}). Consequently  we obtain the following equal $x^+$-time 
commutation relations:
\begin{eqnarray}
&[\tilde{\Sigma}(x),X(y)]=-\frac{i}{\partial_-}\delta^{(3)}
(\vecx^--\vecy^-),\quad [\partial^+\tilde{\Sigma}(x),X(y)]=0,
\label{eq:3.27}&\\ 
&[\tilde{\Sigma}(x),\pi^i(y)]=0,\quad [\partial^+\tilde{\Sigma}(x),
\pi^i(y)]=-\frac{i}{\sqrt{-{\nabla}_{\bot}^{\;\;2}}}
\partial_-\partial_i\delta^{(3)}(\vecx^--\vecy^-), \label{eq:3.28}& \\ 
&[\tilde{\Sigma}(x),\Psi(y)]=-e\frac{(\partial_-)^{-1}}{\sqrt{-{\nabla}_{\bot}
^{\;\;2}}}\delta^{(3)}(\vecx^--\vecy^-)\Psi(y), \quad 
[\tilde{\Sigma}(x),J_-(y)]=0,\label{eq:3.29}& \\
&\hspace*{-4mm}[\partial^+\tilde{\Sigma}(x),\Psi(y)]=\hspace*{-1mm}
\frac{en_-^{-1}}{\sqrt{-{\nabla}_{\bot}
^{\;\;2}}}\delta^{(3)}(\vecx^--\vecy^-)\gamma^+\gamma_-\Psi(y), \;\;
[\partial^+\tilde{\Sigma}(x),J_-(y)]=0,\label{eq:3.30}& \\
&[\tilde{\Sigma}(x),J^+(y)]=[\partial^+\tilde{\Sigma}(x),J^+(y)]=0.&
\end{eqnarray}

Now we can calculate the commutation relations of $C$, which is introduced as 
the integration constant in (\ref{eq:3.7}). We rewrite (\ref{eq:3.7}) to 
express $C$ in terms of operators whose commutation relations are known; we 
get 
\begin{equation}
\frac{\sqrt{-{\nabla}_{\bot}^{\;\;2}}}
{{\nabla}_T^{\;\;2}}C
=\frac{\partial^+}{\partial_-}\tilde{\Sigma}-
\frac{\partial_-^{\;-1}}{\sqrt{-{\nabla}_{\bot}^{\;\;2}}}J_--X  
+\frac{\sqrt{-{\nabla}_{\bot}^{\;\;2}}}{{\nabla}_T^{\;\;2}}
\left(n_-x^-B+\frac{n_-n_+}{{\nabla}_T^{\;\;2}}\partial_+B\right). 
\label{eq:3.32}
\end{equation}
Then using (\ref{eq:3.15}),(\ref{eq:3.16}) and $(\ref{eq:3.24})\sim
(\ref{eq:3.30})$ we obtain at $x^+=y^+$
\begin{eqnarray}
&[\frac{1}{{\nabla}_T^{\;\;2}}C(x),\tilde{\Sigma}
(y)]= [\frac{1}{{\nabla}_T^{\;\;2}}C(x),\partial^+\tilde{\Sigma}(y)]=
[\frac{1}{{\nabla}_T^{\;\;2}}C(x),\Psi(y)]=0,\label{eq:3.33}&\\
&[\frac{1}{{\nabla}_T^{\;\;2}}C(x),X(y)]=[\frac{1}{{\nabla}_T^{\;\;2}}C(x),
J^+(y)]=[\frac{1}{{\nabla}_T^{\;\;2}}C(x),J_-(y)]=0.\label{eq:3.34}&
\end{eqnarray}
If we use (\ref{eq:2.7}) and (\ref{eq:3.26}), then we obtain the following 
four dimensional commutation relation
\begin{eqnarray}
[\frac{1}{{\nabla}_T^{\;\;2}}C(x),B(y)]
=[\frac{-1}{\sqrt{-{\nabla}_{\bot}^{\;\;2}}}X(x),B(y)]
=i\delta^{(3)}(\vecx^+-\vecy^+).  \label{eq:3.35}
\end{eqnarray}
It follows from $(\ref{eq:3.33})\sim(\ref{eq:3.35})$ that
\begin{equation}
[\frac{1}{{\nabla}_T^{\;\;2}}C(x),\frac{1}{{\nabla}_T^{\;\;2}}C(y)]|_{x^+=y^+}
=0.  \label{eq:3.36}
\end{equation}

It is clear that $A_i^{\bot}{\equiv}A_i-\frac{\partial_i
\partial_j}{{\nabla}_{\bot}^{\;\;2}}A_j$ and $\pi^i_{\bot}{\equiv}\pi^i-
\frac{\partial_i\partial_j}{{\nabla}_{\bot}^{\;\;2}}\pi^j$ are gauge invariant 
and thus commute with $B$ and $C$ at equal $x^+$-time. Therefore $B$ and $C$ 
do not give rise to any unwanted contributions to the equal $x^+$-time canonical 
commutation relations. Now that we have 
$A_i^{\bot},\pi^i_{\bot},\tilde{\Sigma},\partial^+\tilde{\Sigma},B$ and $C$ 
as fundamental operators, we can reconstruct our formulation in terms of 
them. To do this, we divide $A_{\mu}$ into physical and RG 
parts as follows
\begin{equation}
A_{\mu}=T_{\mu}
-\frac{n_{\mu}}{{\nabla}_T^{\;\;2}}B+\partial_{\mu}\Lambda \label{eq:3.37}
\end{equation}
where 
\begin{eqnarray}
&T_+=-\frac{\sqrt{-\nabla_{\bot}^{\;\;2}}}{\partial_-}\tilde{\Sigma}
=\frac{1}{\partial_-^{\;2}}(\partial_i\pi^i-J^+),\label{eq:3.38}& \\
&T_-=0, \quad 
T_i= A_i^{\bot}+\frac{1}{\sqrt{-\nabla_{\bot}^{\;\;2}}}
\frac{\partial_i}{\partial_-}\left(\partial^+\tilde{\Sigma}-
\frac{1}{\sqrt{-\nabla_{\bot}^{\;\;2}}}J_-\right), \label{eq:3.39}& \\
&\pi^i=\partial^+A_i^{\bot}+
\frac{\partial_i}{\sqrt{-{\nabla}_T^{\;\;2}}}\left(\partial_-\tilde{\Sigma}
-\frac{1}{\sqrt{-{\nabla}_T^{\;\;2}}}J^+\right).& \label{eq:3.40}
\end{eqnarray}
It is easily seen that $T_{\mu}$ satisfies
\begin{eqnarray}
&\partial^{\mu}f_{\mu\nu}=J_{\nu},\quad (f_{\mu\nu}=\partial_{\mu}T_{\nu}
-\partial_{\nu}T_{\mu}),\label{eq:3.41}&  \\ 
&\partial^{\mu}T_{\mu}=-\frac{1}{\partial_-}J_-,\label{eq:3.42}& 
\end{eqnarray}
and that $T_i$ and $\pi^i$ satisfy the equal $x^+$-time canonical quantization 
conditions
\begin{eqnarray}
&[T_i(x),T_j(y)]=[\pi^i(x),\pi^j(y)]=0, \quad [T_i(x),\pi^j(y)]=i\delta_{ij}
\delta^{(3)}(\vecx^--\vecy^-),& \nonumber \\
&[T_i(x),\Psi(y)]=[\pi^i(x),\Psi(y)]=0.\label{eq:3.43}& 
\end{eqnarray}

\section{Translational generators in the axial gauge formulation}

In this section we start with the  canonical energy-momentum tensor  
\begin{equation}
T^{{\mu}{\nu}}=-F^{\nu\sigma}\partial^{\mu}A_{\sigma}+\frac{g^{\mu
\nu}}{4}F^{\rho\sigma}F_{\rho\sigma}
+i\bar{\Psi}\gamma^{\nu}\partial^{\mu}\Psi \label{eq:4.1}
\end{equation}
and obtain the conserved translation generators. Note that in our 
formulation we 
have to resort to a nonstandard way of deriving the conserved generators. This is because $A_{\mu}$ in (\ref{eq:3.37}) has the term 
depending explicitly on $x^-$ as well as $x^-$-independent terms. As a result, the 
traditional formula for the Hamiltonian contains divergences. This reflects 
the fact that the integral $\int_{-\infty}^{\infty}\partial_-
T_{\mu}^{\;\;-}dx^-$ does not vanish, although $x^-$ is one of space 
coordinates. From the divergence equation, ${\partial }_{\nu }T_
{{\mu}}^{\;\;{\nu }}=0$, we obtain
\begin{equation}
{\oint }T_{{\mu}}^{\;\;{\nu }}d{\sigma }_{\nu }=0 \, ,  \label{eq:4.2}
\end{equation}
\begin{wrapfigure}[13]{r}{60mm}{
\includegraphics[scale=0.65]{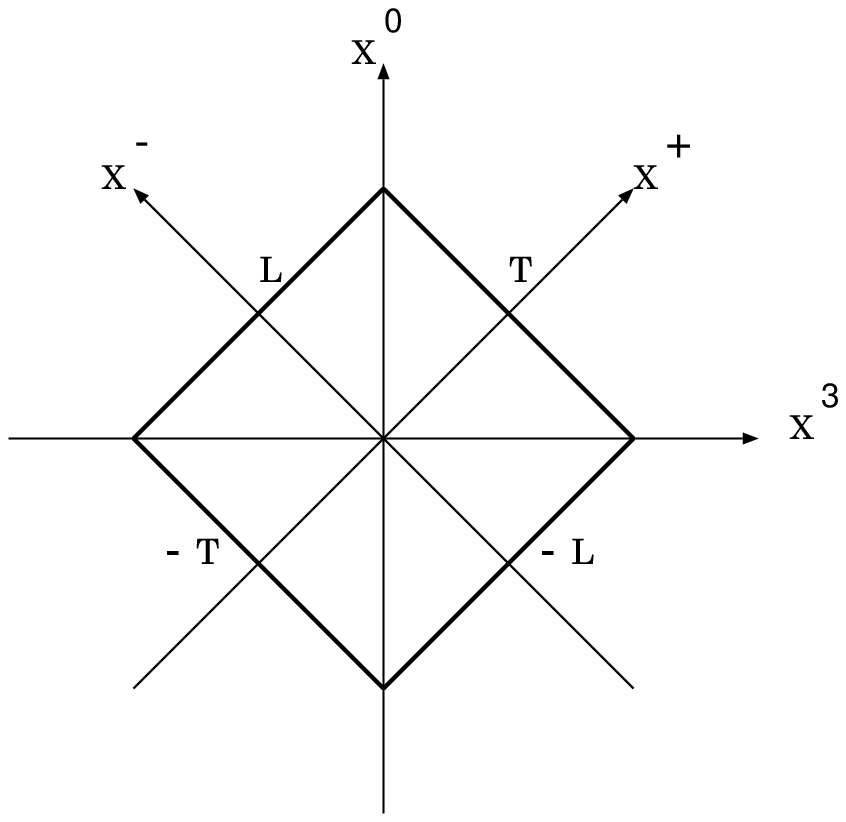}}
\end{wrapfigure}
where the integral is taken over a closed surface.
From this we shall derive the conserved generators.
 For the transverse directions we are justified in assuming that the integral 
${\int}^{\infty}_{-\infty}{\partial}_iT_{\mu}^{\;\;i}dx^i$ vanishes. 
(Here, repeated indices do not imply sum over $i$.) Therefore, as the 
closed surface we use the one shown in Fig.~1, 
whose bounds $T$ and $L$ are taken to ${\infty}$ after the calculations are 
finished. A similar procedure was first used by one of 
the authors of the present paper.\cite{rf:11}
It is straightforward to obtain 
\begin{equation}
0={\int}d^2\vecx_{\bot} \left( {\int }^{L}_{\!\!\!-L}dx^- 
\left[T_{\mu}^{\;\;+}(x) \right]^{x^+=T}_{x^+=-T}+
{\int }_{\!\!\!-T}^{T}dx^+ \left[T_{\mu}^{\;\;-}
(x) \right]^{x^-=L}_{x^-=-L} \right)   \label{eq:4.3}
\end{equation}
where 
\begin{eqnarray}
\left[T_{\mu}^{\;\;+}(x) \right]^{x^+=T}_{x^+=-T}&=&
T_{\mu}^{\;\;+}(x)|_{x^+=T}-T_{\mu}^{\;\;+}(x)|_{x^+=-T}, 
\nonumber \\
\left[T_{\mu}^{\;\;-}(x) \right]^{x^-=L}_{x^-=-L}&=&
T_{\mu}^{\;\;-}(x)|_{x^-=L}-T_{\mu}^{\;\;-}(x)|_{x^-=-L}.  
\label{eq:4.4}
\end{eqnarray}

Let us illustrate the case of $\mu=+$ in detail. Because $A_{\mu}$ has the 
RG term, we divide $T_+^{\;\;+}$ and $T_+^{\;\;-}$ into terms containing 
only physical operators, terms containing
products of physical operators and RG operators and terms consisting solely of RG 
operators. Then, we  
rewrite these terms, which are to be integrated by parts, in the form of 
derivatives minus terms resulting from the integration by parts. 
As a result we obtain
\begin{eqnarray}
&{}&T_+^{\;\;+}
=F^+_{\;\;i}\partial_+A_i-\frac{1}{2}\{(F_{-+})^2+F^+_{\;\;i}F_{+i}
+F^-_{\;\;i}F_{-i}-(F_{12})^2\}+i\bar{\Psi}\gamma^+\partial_+\Psi 
\nonumber \\
&=&\frac{1}{2}\{(f_{-+})^2+f_{+i}f^{\scriptscriptstyle +}_{\;\;i}-f_{-i}
f^{\scriptscriptstyle -}_{\;\;i}+(f_{12})^2
\}+i\bar{\Psi}\gamma^+D_+\Psi-\partial_if^{\scriptscriptstyle +}_{\;i}
\partial_+\Lambda \nonumber \\
&+&J^+(-\frac{n_+}{{\nabla}_T^{\;\;2}}B+\partial_+\Lambda)
-\partial_-\left(f_{-+}T_++T_+\frac{n_-}{{\nabla}_T^{\;\;2}}\partial_+B
+T_i\frac{1}{{\nabla}_T^{\;\;2}}\partial_iB\right) 
\nonumber \\
&+&\partial_i\left(f^{\scriptscriptstyle +}_{\;i}(T_++\partial_+\Lambda)
\right)-\frac{1}{2}\left(\frac{n_-}{{\nabla}_T^{\;\;2}}\partial_+B
\frac{n_-}{{\nabla}_T^{\;\;2}}\partial_+B
+\frac{n_-}{{\nabla}_T^{\;\;2}}\partial_iB\frac{1}
{{\nabla}_T^{\;\;2}}\partial_iB\right)\hspace*{-1mm}, \label{eq:4.5} \\
&{}&T_+^{\;\;-}=F_{-+}\partial_+A_++F^-_{\;\;i}\partial_+A_i
+i\bar{\Psi}\gamma^-\partial_+\Psi 
\nonumber \\
&=&f^-_{\;\;i}f_{+i}+i\bar{\Psi}\gamma^-D_+\Psi + \frac{1}{\partial_-}J^-
\frac{\partial_+}{{\nabla}_T^{\;\;2}}B
-J^-\frac{n_+}{{\nabla}_T^{\;\;2}}B+B\frac{1}{{\nabla}_T^{\;\;2}}
\partial_+C \nonumber \\
&+&\partial_i\left(f^-_{\;\;i}(T_++\partial_+\Lambda)+\partial_+T_i
\frac{1}{{\nabla}_T^{\;\;2}}B+\frac{1}{{\nabla}_T^{\;\;2}}\partial_iB
\frac{n_-n_+}{({\nabla}_T^{\;\;2})^2}\partial_i\partial_+^{\;2}B-
\frac{1}{{\nabla}_T^{\;\;2}}\partial_iB\frac{1}{{\nabla}_T^{\;\;2}}
\partial_+C \right) 
\nonumber \\
&+&\partial_+\left(f_{-+}(T_++\partial_+\Lambda)-\partial_iT_i
\frac{1}{{\nabla}_T^{\;\;2}}B+\frac{n_-}{{\nabla}_T^{\;\;2}}\partial_+B
\frac{n_-n_+}{({\nabla}_T^{\;\;2})^2}\partial_+^{\;3}B
-\frac{1}{{\nabla}_T^{\;\;2}}B\frac{n_-n_+}
{{\nabla}_T^{\;\;2}}\partial_+B \right)   \nonumber \\
&-&\partial_+\left(\frac{n_-}{{\nabla}_T^{\;\;2}}\partial_+B
\frac{1}{{\nabla}_T^{\;\;2}}\partial_+C\right)
+\frac{x^-}{2}\partial_+\left(\frac{n_-}
{{\nabla}_T^{\;\;2}}\partial_+B\frac{n_-}{{\nabla}_T^{\;\;2}}
\partial_+B+\frac{n_-}{{\nabla}_T^{\;\;2}}\partial_iB
\frac{1}{{\nabla}_T^{\;\;2}}\partial_iB \right)\hspace*{-1mm}. \label{eq:4.6}
\end{eqnarray}
When we substitute (\ref{eq:4.5}) and (\ref{eq:4.6}) into (\ref{eq:4.3}) with 
$\mu=+$, the derivative terms with respect to $x^i$ give vanishing 
contributions, whereas the last $x^-$ independent term of (\ref{eq:4.5}) is 
cancelled out by the last linear term of $x^-$ in (\ref{eq:4.6}). We also see 
that $B$ and $C$ contained in the derivative terms with respect to $x^+$ in 
the fourth and fifth lines of (\ref{eq:4.6}) give vanishing contributions in 
the limit $T{\to}\infty$ because $x^+$ is one of spatial coordinates for $B$ 
and $C$. As to $B\frac{1}{{\nabla}_T^{\;\;2}}\partial_+C$ in the second line,
the contribution from the upper bound $L$ cancels out that from the lower 
bound $-L$. However we keep it, because it is the Hamiltonian density for the 
RG fields. We see furthermore, that the derivative term with respect to $x^-$ 
in the third line of (\ref{eq:4.5}) gives rise to a vanishing contribution, 
partly because its first term is cancelled out by the corresponding one in the 
derivative term with respect to $x^+$ in the fourth line of (\ref{eq:4.6}) and 
partly because the physical $T_+$ and $T_i$ vanish in the limit $L{\to}\infty$.
 
With the redundant terms eliminated, we turn to consideration of the 
regularization term. The first term in the third line of (\ref{eq:4.5}) is the 
relevant one. We see that for this to work as the regularization term
we have to get rid of $-\partial_if^{\scriptscriptstyle +}_{\;i}\partial_+
\Lambda $, the remainder of the term $F^+_{\;\;i}\partial_+A_i$, in the second 
line of (\ref{eq:4.5}). This is because if we combine this with $J^+\partial_+
\Lambda$, then we get 
\begin{equation}
(J^+-\partial_if^{\scriptscriptstyle +}_{\;i})\partial_+\Lambda=
-\partial_-f_{-+}\partial_+\Lambda
=\partial_-(T_+\partial_-\partial_+\Lambda-f_{-+}\partial_+\Lambda) 
\label{eq:4.7}
\end{equation}
so that all the derivative terms with respect to $x^-$  in (\ref{eq:4.5}) are 
cancelled out by corresponding derivative terms with respect to $x^+$ in 
(\ref{eq:4.6}). This results in losing the necessary regularizing term. We 
remark that Robertson and McCartor \cite{rf:8} overlooked this fact so that 
they did not obtain the regularization term in the light front formulation of 
QED. We have to discard the unnecessary term in such a way that the divergence 
equation is preserved. It turns out that if we supplement $T_+^{\;\;-}$ with
\begin{equation}
\partial_+\left(\frac{1}{\partial_-^{\;2}}J^+\partial_+\partial_-\Lambda
-\frac{1}{\partial_-}J^+\partial_+\Lambda\right) \label{eq:4.8}
\end{equation}
and if we denote $T_+^{\;\;+}$ with $-\partial_if^{\scriptscriptstyle +}_{\;i}
\partial_+\Lambda $ subtracted as $\theta_+^{\;\;+}$, and $T_+^{\;\;-}$ with 
(\ref{eq:4.8}) added as $\theta_+^{\;\;-}$, then we obtain the following 
divergence equation:
\begin{equation}
\partial_+\theta_+^{\;\;+}+\partial_-\theta_+^{\;\;-}=
\partial_j(f_{+i}f_{ji})-\partial_i(f_{-+}f^-_{\;\;i})-i\partial_i(\bar{\Psi}
\gamma^i\partial_+\Psi)-\partial_i(J_i(T_++\partial_+\Lambda)) \label{eq:4.9} 
\end{equation}
where
\begin{eqnarray}
{\theta}_+^{\;\;+}&=&\frac{1}{2}\{(f_{-+})^2+f_{+i}
f^{\scriptscriptstyle +}_{\;\;i}-f_{-i}f^{\scriptscriptstyle -}_{\;\;i}
+(f_{12})^2\} \nonumber \\
&+&\bar{\Psi}(m-i\gamma^-\partial_--i\gamma^iD_i)\Psi 
+J^+(-\frac{n_+}{{\nabla}_T^{\;\;2}}B+\partial_+\Lambda), \nonumber\\ 
{\theta}_+^{\;\;-}&=&f^-_{\;\;i}f_{+i}-n_-^{\;-1}
\bar{\Psi}\gamma^-\gamma^+(m-i\gamma^-\partial_--i\gamma^iD_i)\Psi \nonumber \\
&+&\frac{1}{\partial_-}J_-\frac{\partial_+}{{\nabla}_T^{\;\;2}}B-J^-\frac{n_+}
{{\nabla}_T^{\;\;2}}B+\partial_+\left(\frac{1}{\partial_-^{\;2}}J^+\partial_+
\partial_-\Lambda-\frac{1}{\partial_-}J^+\partial_+\Lambda\right).
\label{eq:4.10}
\end{eqnarray}
Eq.(\ref{eq:4.9}) indicates that it is enough to consider only 
${\theta}_+^{\;\;+}$ and ${\theta}_+^{\;\;-}$. 

Now taking the limits $L{\to}
{\infty}$ and $T{\to}{\infty}$ in (\ref{eq:4.3}) gives us a conservation law 
which contains RG fields:
\begin{equation}
0={\int}d^3\vecx^- \left[{\theta}_+^{\;\;+} \right]^
{x^+={\infty}}_{x^+=-{\infty}}
+{\int}d^3\vecx^+ \left[{\theta}_+^{\;\;-}+B\frac{1}{{\nabla}_T^{\;\;2}}
\partial_+C \right]^{x^-={\infty}}_{x^-=-{\infty}}.   \label{eq:4.11}
\end{equation}
The derivative term in (\ref{eq:4.8}) gives a vanishing contribution to 
(\ref{eq:4.11}) and  any other terms of ${\theta}_+^{\;\;-}$ are assumed to 
vanish in the limits $x^-{\to}{\pm}\infty$ in accordance with the traditions 
of the AG theories. Then we obtain that the first term is conserved by 
itself. We can add to it the constant Hamiltonian for the RG fields and thus 

we have
\begin{equation}
P_+={\int}\hspace*{-1mm}d^3\vecx^- {\theta}_+^{\;\;+}(x) 
+{\int}\hspace*{-1mm}d^3\vecx^+ B(x)\frac{1}{{\nabla}_T^{\;\;2}}
{\partial}_+C(x). \label{eq:4.12}
\end{equation}

We can similarly derive conservation laws for the $-$ and transverse 
directions as follows:
\begin{eqnarray}
&0={\int}d^3\vecx^- \left[{\theta}_-^{\;\;+} \right]^
{x^+={\infty}}_{x^+=-{\infty}}
+{\int}d^3\vecx^+ \left[{\theta}_-^{\;\;-}-\frac{1}{2}B\frac{n_-}
{{\nabla}_T^{\;\;2}}B \right]^{x^-={\infty}}_{x^-=-{\infty}}, \label{eq:4.13}&
\\ 
&0={\int}d^3\vecx^- \left[{\theta}_i^{\;\;+} \right]^
{x^+={\infty}}_{x^+=-{\infty}}
+{\int}d^3\vecx^+ \left[{\theta}_i^{\;\;-}+B\frac{1}{{\nabla}_T^{\;\;2}}
\partial_iC \right]^{x^-={\infty}}_{x^-=-{\infty}}& \label{eq:4.14}
\end{eqnarray}
where
\begin{eqnarray}
{\theta}_-^{\;\;+}&=&f^+_{\;\;i}f_{-i}+i\bar{\Psi}\gamma^+\partial_-\Psi, 
\label{eq:4.15}\\
{\theta}_-^{\;\;-}&=&\frac{1}{2}\{(f_{-+})^2-f^{\scriptscriptstyle +}_{\;\;i}
f_{+i}+f^{\scriptscriptstyle -}_{\;\;i}f_{-i}+(f_{12})^2\} 
+i\bar{\Psi}\gamma^-\partial_-\Psi \nonumber \\
&-&J^-\frac{n_-}{{\nabla}_T^{\;\;2}}B
-\partial_+(\frac{1}{\partial_-}J^+\frac{n_-}{{\nabla}_T^{\;\;2}}B), \\
{\theta}_i^{\;\;+}&=&f^+_{\;\;j}\partial_iT_j
+i\bar{\Psi}\gamma^+\partial_i\Psi, \label{eq:4.17} \\
{\theta}_i^{\;\;-}&=&f_{-+}\partial_i+f^-_{\;\;j}\partial_iT_j
+i\bar{\Psi}\gamma^-\partial_i\Psi \nonumber \\
&-&\frac{\partial_i}{\partial_-}J_-\frac{1}{{\nabla}_T^{\;\;2}}B
-J^-\partial_i\Lambda+\partial_+\left(\frac{1}{\partial_-^{\;2}}J^+\partial_i
\partial_-\Lambda-\frac{1}{\partial_-}J^+\partial_i\Lambda\right).
\end{eqnarray}
From (\ref{eq:4.13}) and (\ref{eq:4.14}) we obtain the generators of 
translations for the $-$ and transverse directions:
\begin{eqnarray}
P_-&=&{\int}d^3\vecx^- {\theta}_-^{\;\;+}(x)  
-\frac{1}{2}{\int}d^3\vecx^+ B(x)\frac{n_-}{{\nabla}_T^{\;\;2}}B(x), 
\label{eq:4.19}\\ 
P_i&=&{\int}d^3\vecx^- {\theta}_i^{\;\;+}(x)  
+{\int}d^3\vecx^+ B(x)\frac{1}{{\nabla}_T^{\;\;2}}\partial_iC(x). 
\label{eq:4.20}
\end{eqnarray}

We end this section by showing that the Heisenberg equations for the 
constituent fields hold. For that purpose we use the expressions 
\begin{equation}
f_{-+}=\frac{1}{\partial_-}(\partial_i\pi^i-J^+), \quad
f^+_{\;\;i}=\pi^i, \quad f_{+i}=\frac{\pi^i-n_+f_{-i}}{(-n_-)}, \quad
f^-_{\;\;i}=\frac{n_+\pi^i-f_{-i}}{(-n_-)} \, ,
\end{equation}
to express the Hamiltonian in terms of $T_i$ and $\pi^i$ as follows
\begin{eqnarray}
&{}&P_+={\int}\hspace*{-1mm}d^3\vecx^- {\theta}_+^{\;\;+}(x) 
+{\int}\hspace*{-1mm}d^3\vecx^+ B(x)\frac{1}{{\nabla}_T^{\;\;2}}
{\partial}_+C(x) \nonumber\\ 
&=&\frac{1}{2}{\int}\hspace*{-1mm}d^3\vecx^-\{(\partial_-^{\;-1}
(\partial_i\pi^i-J^+))^2+(-n_-)^{-1}(\pi^i-n_+f_{-i})^2 +(-n_-)(f_{-i})^2
+(f_{12})^2 \} \nonumber \\
&+&\hspace*{-1mm}{\int}\hspace*{-1mm}d^3\vecx^-\{\bar{\Psi}
(m-i\gamma^-\partial_--i\gamma^iD_i)\Psi+\hspace*{-1mm}
J^+(\partial_+\Lambda-\frac{n_+}{{\nabla}_T^{\;\;2}}B) \}
+\hspace*{-1mm}{\int}\hspace*{-1mm}d^3\vecx^+  B\frac{\partial_+}
{{\nabla}_T^{\;\;2}}C. \label{eq:4.22}
\end{eqnarray}
Note here that we can use the equal $x^+$-time commutation relations to 
calculate the commutation relations with the operators contained in 
${\theta}_+^{\;\;+}(x)$, while $B$ and $C$ are regarded as independent of 
other constituent operators so that they commute with the others at all times. 
As a consequence we obtain
\begin{eqnarray}
&[T_i(x),P_+]=i\frac{\partial_i}{\partial_-^{\;2}}(\partial_j\pi^j-J^+)
+i\frac{\pi^i-n_+f_{-i}}{(-n_-)}=i(\partial_iT_++f_{+i})=i\partial_+T_i,
\label{eq:4.23}& \\
&[\Psi(x),P_+]=e(T_+-\frac{n_+}{{\nabla}_T^{\;\;2}}B+\partial_+\Lambda)\Psi
-\frac{\gamma^+}{n_-}(m-i\gamma^-\partial_--i\gamma^iD_i)\Psi(x)&
\label{eq:4.24}
\end{eqnarray}
where we have used the equality $(\gamma^+)^2=-n_-$ as well as the fact that 
$\Psi$ commutes with $B$ and $C$ contained in ${\theta}_+^{\;\;+}$ at equal 
$x^+$-time because they are multiplied by $\frac{1}{{\nabla}_T^{\;\;2}}$. 
Equating (\ref{eq:4.24}) with $\partial_+\Psi$ and then multiplying both 
sides by $\gamma^+$ provides us with the field equation of $\Psi$.

In quite the same manner we obtain 
\begin{equation}
[T_i(x),P_r]=i\partial_rT_i, \quad [\Psi(x),P_r]=i\partial_r\Psi. 
\quad (r=-,1,2)
\end{equation}
Equations for $B$ and $C$ are calculated by the same rules and we get
\begin{eqnarray}
&[\frac{1}{{\nabla}_T^{\;\;2}}B(x),P_r]=i\frac{\partial_r}{{\nabla}_T^{\;\;2}}B
, \quad [\frac{1}{{\nabla}_T^{\;\;2}}C(x),P_r]=i\frac{\partial_r}
{{\nabla}_T^{\;\;2}}C, \quad (r=+,1,2)& \\
&[\frac{1}{{\nabla}_T^{\;\;2}}B(x),P_-]=0, \quad 
[\frac{1}{{\nabla}_T^{\;\;2}}C(x),P_-]=-i\frac{n_-}{{\nabla}_T^{\;\;2}}B.&
\end{eqnarray}
It follows  that
\begin{equation}
[A_i(x),P_{\mu}]=i\partial_{\mu}A_i.
\end{equation}

\section{A perturbation formulation free from infrared divergences}

\subsection{ calculation of the propagator}

Now it is straightforward to develop an $x^+$-time ordered perturbation theory 
by employing the following free and interaction Hamiltonians:
\begin{eqnarray}
H_0
&=&\frac{1}{2}{\int}\hspace*{-1mm}d^3\vecx^-\{(\partial_-^{\;-1}
\partial_i\pi^i)^2-n_-^{\;-1}(\pi^i-n_+f_{-i})^2 -n_-(f_{-i})^2
+(f_{12})^2 \} \nonumber \\
&+&\hspace*{-1mm}{\int}\hspace*{-1mm}d^3\vecx^-\bar{\Psi}
(m-i\gamma^-\partial_--i\gamma^i\partial_i)\Psi
+\hspace*{-1mm}{\int}\hspace*{-1mm}d^3\vecx^+  B\frac{\partial_+}
{{\nabla}_T^{\;\;2}}C, \\
H_{I}&=& {\int}\hspace*{-1mm}d^3\vecx^- \{J^{\mu}A_{\mu}-\frac{1}{2}J^+\frac{1}
{\partial_-^{\;2}}J^+ \} \label{eq:5.1}
\end{eqnarray}
where  we have denoted the free fields in the interaction representation 
with the same notation as those in the Heisenberg representation. Thus 
\begin{equation}
A_{\mu}=T_{\mu}+\Gamma_{\mu},\quad \Gamma_{\mu}=-\frac{n_{\mu}}
{{\nabla}_T^{\;\;2}}B+\partial_{\mu}\Lambda
\end{equation}
is the free gauge field. The physical part, $T_{\mu}=a_{\mu}
-\frac{\partial_{\mu}}{\partial_-}a_-$, is described in terms of free fields 
$\partial_+a_--\partial_-a_+=\sqrt{-{\nabla}_{\bot}^{\;\;2}}\tilde{\Sigma}$ 
and $a_i^{\bot}=a_i-\frac{\partial_i\partial_j}{{\nabla}_{\bot}^{\;\;2}}a_j$ 
as 
\begin{eqnarray}
&T_+=a_+-\frac{\partial_+}{\partial_-}a_-
=-\frac{\sqrt{-{\nabla}_{\bot}^{\;\;2}}}{\partial_-}\tilde{\Sigma},& \\
&T_-=0, \quad 
T_i=a_i-\frac{\partial_i}{\partial_-}a_-=a_i^{\bot}+\frac{\partial_i}
{\sqrt{-{\nabla}_{\bot}^{\;\;2}}}\frac{\partial^+}{\partial_-}\tilde{\Sigma}.&
\end{eqnarray}
If we consider the fact that the conjugate momentum of $T_i$ is given by 
\begin{equation}
\pi^i=f^+_{\;\;i}=\partial^+a_i^{\bot}+\frac{\partial_i}
{\sqrt{-{\nabla}_{\bot}^{\;\;2}}}\partial_-\tilde{\Sigma},
\end{equation}
then we see that $\tilde{\Sigma},\;\partial^+\tilde{\Sigma}$ and $a_i^{\bot},
\;\partial^+a_i^{\bot}$ are two pairs of canonical variables. As a consequence
we can express $T_{\mu}$ in terms of creation and annihilation operators 
as  
\begin{equation}
T_{\mu}(x)=\frac{1}{\sqrt{2(2{\pi})^3}}\int \frac{d^3\veck_-}{\sqrt{k^+}}
\sum_{{\lambda}=1}^2{\epsilon}_{\mu}^{({\lambda})}(k) \{
a_{\lambda}(\veck_-)e^{-ik\cdot x}+ {\rm h.c.} \} \label{eq:5.7} 
\end{equation}
where
\begin{equation}
  k^+=\sqrt{k_-^2-n_-k_{\bot}^2},  
\quad k^-=\frac{k_--n_+k^+}{n_-}, \quad k_+=\frac{n_+k_--k^+}{n_-}, 
\quad k_{\bot}=\sqrt{k_1^2+k_2^2} 
\end{equation}
and the operators $a_{{\lambda}}(\veck_-)$ and $a_{\lambda}^{\dagger}
(\veck_-)\;({\lambda}=1,2)$ are normalized so as to satisfy the usual 
commutation relations,
\begin{equation}
[a_{\lambda}(\veck_-),\;a_{{\lambda}^{\prime}}(\vecq_-)]=0, \quad
[a_{\lambda}(\veck_-),\;a_{{\lambda}^{\prime}}^{\dagger}(\vecq_-)]=
{\delta}_{{\lambda}{\lambda}^{\prime}}{\delta}^{(3)}(\veck_--\vecq_-).
\end{equation}
Note that
\begin{eqnarray}
&{\epsilon}_{\mu}^{(1)}(k)=\left( -\frac{k_{\bot}}{k_-},\;0,\;-\frac{k^+k_1}
{k_-k_{\bot}},\;-\frac{k^+k_2}{k_-k_{\bot}}\right),& \\ 
&{\epsilon}_{\mu}^{(2)}(k)=(0,\;0,\;-\frac{k_2}{k_{\bot}},\;
\frac{k_1}{k_{\bot}})&
\end{eqnarray} 
are the polarization vectors satisfying
\begin{eqnarray}
&k^{\mu}{\epsilon}_{\mu}^{({\lambda})}(k)=0, \quad 
n^{\mu}{\epsilon}_{\mu}^{({\lambda})}(k)=0, \quad ({\lambda}=1,2)& \\  
&\sum_{{\lambda}=1}^2{\epsilon}_{\mu}^{({\lambda})}(k)
{\epsilon}_{\nu}^{({\lambda})}(k)=-g_{{\mu}{\nu}}+\frac{n_{\mu}k_{\nu}+
n_{\nu}k_{\mu}}{k_-}-n^2\frac{k_{\mu}k_{\nu}}{k_-^2}.& 
\end{eqnarray}

We expand $B$ and $C$ in terms of zero-norm creation and annihilation 
operators  as follows:
\begin{equation}
-\frac{1}{{\nabla}_T^{\;\;2}}B(x) 
=\frac{1}{\sqrt{(2{\pi})^3}}\int \frac{d^3\veck_+}{\sqrt{k_+}}
{\theta}(k_+) \{ B(\veck_+)e^{-ik\cdot x}+ 
B^{\dagger}(\veck_+)e^{ik\cdot x}\}|_{x^-=0},  
\end{equation}  
\begin{equation}
C(x) =\frac{i}{\sqrt{(2{\pi})^3}}\int d^3\veck_+{\sqrt{k_+}}
 {\theta}(k_+)\{ C(\veck_+)e^{-ik\cdot x}- 
C^{\dagger}(\veck_+)e^{ik\cdot x}\}|_{x^-=0},  
\end{equation} 
where
\begin{equation}
[B(\veck_+),\;C^{\dagger}(\vecq_+)]=[C(\veck_+),\;B^{\dagger}(\vecq_+)]=
-{\delta}^{(3)}(\veck_+-\vecq_+),
\end{equation}
and all other commutators are zero. We note here that limiting the $k_+$-
integration region to be $(0,{\infty})$ is indispensable to obtain the 
ML form of gauge field propagator. 
We define the vacuum state and physical space $V_P$, respectively by
\begin{equation}
B(\veck_+)|{\rm{\Omega}}{\rangle}=C(\veck_+)|{\rm{\Omega}}{\rangle}=0,  
\end{equation}
\begin {equation}
  V_P=\{\; |{\rm phys}{\rangle}\; | \;B(\veck_+)|{\rm phys}{\rangle}=0\; \}. 
\end{equation}

Now we can calculate the $x^+$-ordered gauge field propagator
\begin{eqnarray}
D_{{\mu}{\nu}}(x-y)&=&{\langle}{\rm{\Omega}}|\{{\theta}(x^+-y^+)A_{\mu}(x)
A_{\nu}(y)+{\theta}(y^+-x^+)A_{\nu}(y)A_{\mu}(x) \}|{\rm{\Omega}}{\rangle}
 \nonumber \\
&=&\frac{1}{(2{\pi})^4}\int d^4qD_{{\mu}{\nu}}(q)e^{-iq \cdot (x-y)}.
\label{eq:5.19}
\end{eqnarray}
It is straightforward to show that its physical part is given by
\begin{equation}
D^p_{{\mu}{\nu}}(q)=\frac{i}{q^2+i{\epsilon}} \left( -g_{{\mu}{\nu}}
+\frac{n_{\mu}q_{\nu}+n_{\nu}q_{\mu}}{q_-}-n^2\frac{q_{\mu}q_{\nu}}{q_-^2 } 
\right)-{\delta}_{{\mu}+}{\delta}_{{\nu}+}\frac{i}{q_-^2},\label{eq:5.20}  
\end{equation}
where $q^2=n_-q_-^2+2n_+q_+q_--n_-q_+^2-q_{\bot}^2$.
We investigate in detail how the RG fields  play roles as regulators.  
In the case that  ${\mu}=i$ and ${\nu}=j$ we obtain the RG contribution 
\begin{equation}
{\langle}{\rm{\Omega}}|T\left( {\Gamma}_i(x){\Gamma}_j(y) \right)
|{\rm{\Omega}}{\rangle}
=\frac{1}{(2{\pi})^4}\int d^4qD^g_{ij}(q)e^{-iq \cdot (x-y)}
\end{equation}
where
\begin{eqnarray}
&D^g_{ij}(q)=q_iq_j{\int}^{\infty}_0dk_+\Bigl[ {\delta}^{\prime}(q_-)
\frac{n_-}{n_-k_+^2+q_{\bot}^2}\Bigr.\left(\frac{i}{k_+-q_+-i{\epsilon}}
-\frac{i}{k_++q_+-i{\epsilon}}\right) & \nonumber \\
&- {\delta}(q_-)\frac{2n_-n_+k_+}{(n_-k_+^2+q_{\bot}^2)^2}
\left(\frac{i}{k_+-q_+-i{\epsilon}}+\frac{i}{k_++q_+-i{\epsilon}}\right) 
\Bigl.\Bigr].& \label{eq:5.22}
\end{eqnarray}
Note that the explicit $x^-$ dependence gives rise to the factor ${\delta}^
{\prime}(q_-)$. Note also that there is no on-mass-shell condition among the RG 
field's momenta $k_+,k_1, k_2$ so that there remains a $k_+$-integration. As a 
consequence there arise singularities resulting from the inverse of the 
hyperbolic Laplace operator. Nevertheless, when we regularize the 
singularities as the principal values, the integral on the first line of 
(\ref{eq:5.22}) turns out to be well-defined. In fact we can rewrite its 
integrand as a sum of simple poles and make use of the integral formulas
\begin{eqnarray}
&{\int}^{\infty}_0dk_+\left( \frac{i}{k_+-q_+-i{\epsilon}}
-\frac{i}{k_++q_+-i{\epsilon}} \right)=-{\pi}{\rm sgn}(q_+), \label{eq:5.23}&\\
&{\rm P}{\int}^{\infty}_0dk_+\left( \frac{1}{k_+-a}-\frac{1}{k_++a} \right)=0,&
\label{eq:5.24}
\end{eqnarray}
where $a=\frac{q_{\bot}}{\sqrt{-n_-}}$. As a result we obtain
\begin{equation}
{\int}^{\infty}_0dk_+
\frac{n_-}{n_-k_+^2+q_{\bot}^2}\left(\frac{i}{k_+-q_+-i{\epsilon}}
-\frac{i}{k_++q_+-i{\epsilon}}\right)=-
\frac{n_-{\pi}{\rm sgn}(q_+)}{n_-q_+^2+q_{\bot}^2}. \label{eq:5.25}
\end{equation}

On the other hand, the integral on the second line of (\ref{eq:5.22}) yields a 
linear divergence, as is seen by rewriting its integrand as a sum of simple 
and double poles:
\begin{eqnarray}
&{}& \frac{2n_+n_-k_+}{(n_-k_+^2+q_{\bot}^2)^2}\left(\frac{i}
{k_+-q_+-i{\epsilon}}+\frac{i}{k_++q_+-i{\epsilon}}\right) \nonumber \\
&=&\frac{2n_+n_-q_+}{(n_-q_+^2+q_{\bot}^2)^2} 
\left(\frac{i}{k_+-q_+-i{\epsilon}}-\frac{i}{k_++q_+-i{\epsilon}}\right) 
-\frac{n_+}{a}\frac{n_-q_+^2-q_{\bot}^2}{(n_-q_+^2+q_{\bot}^2)^2} 
\left(\frac{i}{k_+-a} \right.\nonumber \\
&-& \left. \frac{i}{k_++a}\right) 
-\frac{n_+}{n_-q_+^2+q_{\bot}^2}\left( \frac{i}{(k_+-a)^2}+\frac{i}{(k_++a)^2}
\right). \label{eq:5.26}
\end{eqnarray}
The integration of the first and second terms on the right hand side can 
be carried out with the help of (\ref{eq:5.23}) and (\ref{eq:5.24}). However we 
cannot regularize a linear divergence resulting from a double pole by the 
principal value prescription.\cite{rf:15} We show below that this linear 
divergence is necessary 
to cancel a corresponding one that occurs in the physical part. For later convenience we 
rewrite the linearly diverging integration in the form
\begin{equation}
{\rm P}{\int}^{\infty}_0dk_+ \left( \frac{1}{(k_+-a)^2}+\frac{1}{(k_++a)^2} 
\right)={\rm P}{\int}^{\infty}_0dk_+\frac{2n_-(n_-k_+^2-q_{\bot}^2)}
{(n_-k_+^2+q_{\bot}^2)^2}.
\label{eq:5.27}
\end{equation}
Substituting  these results and  (\ref{eq:5.27}) into  (\ref{eq:5.22}) 
yields
\begin{equation}
D^g_{ij}(q)=\frac{n_-q_iq_j}{q^2+i{\epsilon}}{\delta}^{\prime}(q_-)
{\pi}{\rm sgn}(q_+)-i\frac{n_+q_iq_j}{q^2+i{\epsilon}}{\delta}(q_-)
{\int}^{\infty}_0dk_+\frac{2n_-(n_-k_+^2-q_{\bot}^2)}
{(n_-k_+^2+q_{\bot}^2)^2}, \label{eq:5.28}
\end{equation}
where we have made use of the identity
\begin{equation}
\frac{1}{q^2+i{\epsilon}}{\delta}^{\prime}(q_-)=
-\frac{1}{n^2q_+^2+q_{\bot}^2}{\delta}^{\prime}(q_-)
+\frac{2n_+q_+}{(n^2q_+^2+q_{\bot}^2)^2}{\delta}(q_-). \label{eq:5.29}
\end{equation}
Thus, for the sum of (\ref{eq:5.20}) and (\ref{eq:5.28}) we obtain
\begin{eqnarray}
&D_{ij}(q)=\frac{i}{q^2+i{\epsilon}}\Bigl( \Bigr.
-g_{ij}-n^2\frac{q_iq_j}{q_-^2}-in_-q_iq_j{\pi}{\rm sgn}(q_+)
{\delta}^{\prime}(q_-) &  \nonumber \\
&\left.-n_+q_iq_j{\delta}(q_-){\int}^{\infty}_0dk_+
\frac{2n_-(n_-k_+^2-q_{\bot}^2)}{(n_-k_+^2+q_{\bot}^2)^2} 
\right).&  \label{eq:5.30}
\end{eqnarray}

Now we can demonstrate that the linear divergence resulting from 
$\frac{1}{q_-^2}$ is canceled by the final term of (\ref{eq:5.30}) when we 
restore $D_{ij}(x)$ by substituting  (\ref{eq:5.30}) into (\ref{eq:5.19}). 
 It should be noted here that $q_-$ is conjugate to the spatial variable $x^-$,
 while $k_+$ is conjugate to the temporal variable $x^+$. To make the infrared 
divergence cancellation mechanism work, both integration variables have to be 
either spatial or temporal. We show in Appendix A that if we change the 
integration variable from the spatial $q_-$ to the temporal 
$k_+=\frac{n_+q_--\sqrt{q_-^2-n_-q_{\bot}^{\;2}}}{n_-}$,
then the following integral formula holds:
\begin{equation}
{\int}^{\infty}_{-\infty}dq_-\frac{1}{q_-^2}+{\int}^{\infty}_0dk_+
\frac{2n_+(n_-k_+^2-q_{\bot}^2)}
{(n_-k_+^2+q_{\bot}^2)^2}=0. \label{eq:5.31}
\end{equation}
From (\ref{eq:5.31}) we see that the linearly diverging terms of the inverse 
Fourier transform of (\ref{eq:5.30}) are given by (we omit $-in^2q_iq_j$ 
for the moment)
\begin{eqnarray}
&{}&{\int}^{\infty}_{-\infty}dq_-\frac{1}{q^2+i{\epsilon}}
\left( \frac{1}{q_-^2}
-{\delta}(q_-){\int}^{\infty}_{-\infty}dq_-\frac{1}{q_-^2}\right)e^{-iq_-x^-} 
\nonumber \\
&=&{\int}^{\infty}_{-\infty}dq_-\frac{e^{-iq_-x^-}}{q_-^2(q^2+i{\epsilon})}+
\frac{1}{n^2q_+^2+q_{\bot}^2}{\int}^{\infty}_{-\infty}dq_-\frac{1}{q_-^2}.  
\nonumber \\
&=&{\int}^{\infty}_{-\infty}dq_-\left(\frac{e^{-iq_-x^-}-1}{q_-^2}
\frac{1}{q^2+i{\epsilon}}+\frac{2n_+q_++n^2q_-}{q_-(n^2q_+^2+q_{\bot}^2)
(q^2+i{\epsilon})}\right).  \label{eq:5.32}
\end{eqnarray}
We see from this that the last integrals diverge at most logarithmically, but 
logarithmic divergences can be regularized by the principal value prescription 
so that there arise no divergences from (\ref{eq:5.32}). This verifies that 
the following identity holds:
\begin{eqnarray}
&\frac{1}{q_-^2}+i{\pi}{\rm sgn}(q_+){\delta}^{\prime}(q_-)
-{\delta}(q_-){\int}^{\infty}_{-\infty}dq_-\frac{1}{q_-^2}& \nonumber \\
&={\rm Pf}\frac{1}{q_-^2}+i{\pi}{\rm sgn}(q_+){\delta}^{\prime}(q_-)
=\frac{1}{(q_-+i{\epsilon}{\rm sgn}(q_+))^2}&  \label{eq:5.33}
\end{eqnarray}
where Pf denotes Hadamard's finite part. Now substituting (\ref{eq:5.33}) into 
(\ref{eq:5.30}) yields the ML form of gauge field propagator:
\begin{equation}
D_{ij}(q)=\frac{i}{q^2+i{\epsilon}}\left(
-g_{ij}-\frac{n^2q_iq_j}{(q_-+i{\epsilon}{\rm sgn}(q_+))^2} \right).
 \label{eq:5.34}
\end{equation}

For other cases we omit detailed demonstrations because the calculations are 
similar. In the case that ${\mu}=+$ and ${\nu}=i$, we obtain the following 
RG contribution
\begin{eqnarray}
&D^g_{+i}(q)=q_i{\int}^{\infty}_0dk_+\Bigl[ \Bigr.{\delta}(q_-)\frac{n_+}
{n_-k_+^2+q_{\bot}^2}\left( \frac{i}{k_+-q_+-i{\epsilon}}-\frac{i}
{k_++q_+-i{\epsilon}} \right) & \nonumber \\
&+{\delta}^{\prime}(q_-)\frac{n_-k_+}{n_-k_+^2+q_{\bot}^2} \left( \frac{i}
{k_+-q_+-i{\epsilon}}+\frac{i}{k_++q_+-i{\epsilon}} \right)& \nonumber \\
&- {\delta}(q_-) \frac{2n_-n_+k_+^2}{(n_-k_+^2+q_{\bot}^2)^2}
\left( \frac{i}{k_+-q_+-i{\epsilon}}-\frac{i}{k_++q_+-i{\epsilon}} \right)
\Bigl.\Bigr]. \label{eq:5.35}
\end{eqnarray}
Because the integral on the first line gives rise to the imaginary part of 
$\frac{1}{q_-+i{\epsilon}{\rm sgn}(q_+)}$, we obtain
\begin{equation}
D_{+i}(q)=\frac{i}{q^2+i{\epsilon}}\left(
\frac{n_+q_i}{q_-+i{\epsilon}{\rm sgn}(q_+)}-\frac{n^2q_+q_i}{(q_-+i{\epsilon}
{\rm sgn}(q_+))^2} \right).
 \label{eq:5.36}
\end{equation}
For the case that ${\mu}={\nu}=+$ the RG contribution turns out to be 
\begin{eqnarray}
&D^g_{++}(q)={\int}^{\infty}_0dk_+\Bigl[ \Bigr. 
{\delta}^{\prime}(q_-)\frac{n_-k_+^2}{n_-k_+^2+q_{\bot}^2}
\Bigl(\frac{i}{k_+-q_+-i{\epsilon}}-\frac{i}{k_++q_+-i{\epsilon}}\Bigr) &
\nonumber \\
&+ {\delta}(q_-) \frac{2n_+k_+q_{\bot}^2}{(n_-k_+^2+q_{\bot}^2)^2}
\left( \frac{i}{k_+-q_+-i{\epsilon}}+\frac{i}{k_++q_+-i{\epsilon}} \right)
\Bigl.\Bigr], \label{eq:5.37}
\end{eqnarray}
so that we obtain
\begin{eqnarray}
&D^g_{++}(q)=\frac{1}{q^2+i{\epsilon}}\Bigl( 2n_+q_+{\delta}(q_-){\pi}{\rm sgn}
(q_+)\Bigr.& \nonumber \\
&\Bigl.+n^2q_+^2{\delta}^{\prime}(q_-){\pi}{\rm sgn}(q_+)-q_{\bot}^2{\delta}
(q_-){\int}^{\infty}_{-\infty}dq_-\frac{i}{q_-^2} \Bigr).
\label{eq:5.38}
\end{eqnarray}
Here it is noteworthy that the last term in (\ref{eq:5.38}) also cancels the 
linear divergence resulting from the contact term in (\ref{eq:5.20}).
In fact it also yields a contact term, as is seen from
$-{\delta}(q_-)\frac{q_{\bot}^2}{q^2+i{\epsilon}}={\delta}(q_-)\left(
1+\frac{n^2q_+^2}{q^2+i{\epsilon}} \right).$ 
Therefore, combining it with the corresponding one in (\ref{eq:5.20}) and 
carrying out the inverse Fourier transform, we obtain 
\begin{equation}
\frac{1}{2{\pi}}{\int}^{\infty}_{-\infty}dq_-\left({\delta}(q_-)
{\int}^{\infty}_{-\infty}dq_-\frac{1}{q_-^2}-\frac{1}{q_-^2}\right)
e^{-q_-x^-}=\frac{1}{{\pi}}{\int}^{\infty}_0dq_-\frac{1-{\rm cos}q_-x^-}{q_-^2}
=\frac{|x^-|}{2}. \label{eq:5.39}
\end{equation}
Consequently, taking into account the fact that the Fourier transform of 
$\frac{|x^-|}{2}$ is $-\frac{1}{2}(\frac{1}{(q_-+i{\epsilon})^2}
+\frac{1}{(q_--i{\epsilon})^2})$,  we obtain
\begin{eqnarray}
&D_{++}(q)=\frac{i}{q^2+i{\epsilon}}\left(-g_{++}+
\frac{2n_+q_+}{q_-+i{\epsilon}{\rm sgn}(q_+)}-\frac{n^2q_+^2}{(q_-+i{\epsilon}
{\rm sgn}(q_+))^2} \right)& \nonumber \\ 
&-{\delta}_{{\mu}+}{\delta}_{{\nu}+}\frac{i}{2}
\left(\frac{1}{(q_-+i{\epsilon})^2}+\frac{1}{(q_--i{\epsilon})^2}\right).&
 \label{eq:5.40}
\end{eqnarray}
This finishes the demonstration that, due to the RG fields, the linear 
divergences are eliminated even in the most singular component of 
$x^+$-ordered  gauge field propagator.

We end this subsection by investigating the light-front limit $\theta{\to}
\frac{\pi}{4}+0.$ On the mass-shell, 
$k^2=0$, $k_+$ is given by $k_+=\frac{n_+k_--\sqrt{k_-^2-n_-k_{\bot}^2}}
{n_-}$ so we get 
\begin{equation}
\lim_{\theta{\to}\frac{\pi}{4}+0}k_+=\left\{\begin{array}{ll}
\frac{k_{\bot}^2}{2k_-} & (k_->0) \\
\infty & (k_-<0)
\end{array}\right.
\end{equation}
and contributions to $T_{\mu}$ in (\ref{eq:5.7}) from the integration 
region in which $k_-<0$ vanish due to the Riemann-Lebesgue lemma. As a result, 
the integration region $k_-$ in $T_{\mu}$ is limited to $(0,\infty)$. 
It is worthwhile noting that in the limit, $n_-=n^2
{\to}0$, so we do not have the linear divergences except for the contact 
term in the most singular component of the gauge field propagator:
\begin{equation}
D_{++}(q)=\frac{2q_+}{q_-}\frac{i}{q^2+i{\epsilon}}-\frac{i}{q_-^2}
-i\frac{2q_+}{q_{\bot}^2}\delta(q_-)(-i\pi{\rm sgn}(q_+))+
\frac{4i}{q_{\bot}^2}\delta(q_-)\int_0^{\infty}dk_+.
\end{equation}
 However, the 
linear divergence resulting from the contact term is also cancelled out due 
to an equality
\begin{equation}
{\int}_{\!\!\!-\infty}^{\infty}dq_-\frac{1}{q_-^2}-{\int}^{\infty}_{\!0}dk_+
\frac{4}{q_{\bot}^2}=0,
\end{equation}
which is verified by changing the integration variable from 
$q_-$ to $k_+=\frac{q_{\bot}^2}{2q_-}$\cite{rf:13}. 

In the light-front formulation $\psi_-=\frac{1}{\sqrt{2}}
\gamma^0\gamma^-\psi$ is a dependent field and  has to be expressed in 
terms of the independent field $\psi_+=\frac{1}{\sqrt{2}}\gamma^0\gamma^+\psi$. 
We follow Morara and Soldati\cite{rf:7} to solve this problem and obtain
the following free and interaction Hamiltonians 
\begin{eqnarray}
H_0
&=&{\int}\hspace*{-1mm}d^3\vecx^-\{\frac{1}{2}(f_{-+})^2+\frac{1}{2}(f_{12})^2 
+i\bar{\psi}\gamma^-\partial_-\psi\} 
+\hspace*{-1mm}{\int}\hspace*{-1mm}d^3\vecx^+  B\frac{\partial_+}
{{\nabla}_{\perp}^{\;\;2}}C, \\
H_{I}&=& {\int}\hspace*{-1mm}d^3\vecx^- \{J^{\mu}A_{\mu}
+e^2\bar{\psi}\gamma^{\mu}A_{\mu}\frac{\gamma^+}{2i\partial_-}
\gamma^{\nu}A_{\nu}\psi-\frac{1}{2}J^+\frac{1}
{\partial_-^{\;2}}J^+ \}. \label{eq:5.45}
\end{eqnarray}
The $x^+$-ordered electron propagator is given by Chang and Yan\cite{rf:16} to 
be 
\begin{equation}
S_F = \frac{i}{(2 \pi)^4} \int d^4p \;\;e^{-i p \cdot x} \left[\frac{(\not\!{p} + m)}{p^2 - m^2 + i \varepsilon} - \ha \frac{\gamma^+}{p^+}\right] \label{fp} .
\end{equation}
We see here that there appears a noncovariant instant term, but its 
contributions to S matrices are cancelled by those resulting from the 
noncovariant fourpoint interaction term. 

\subsection{ calculation of the electron self energy}

In order to make comparison with previous calculations we shall regulate our 
calculation with the inclusion of Pauli-Villars fields. If we perform the 
calculation in the usual order for equal-time quantization, we should expect 
to need only one Pauli-Villars photon. However, it is common experience that 
more Pauli-Villars fields are needed when one performs the $x^+$ integral 
first. In Feynman gauge three Pauli-Villars photons\cite{rf:10a} or one 
Pauli-Villars electron and one Pauli-Villars photon are needed.\cite{rf:10b}  
In light-front gauge with higher derivatives so that $A_+$ is a degree of 
freedom, a much more complicated regulation is required including (in addition 
to the higher derivatives): three Pauli-Villars electrons, a Pauli-Villars 
photon and two carefully treated cutoffs\cite{rf:10b}. In all these cases the 
Pauli-Villars electrons are included with flavor changing currents that break 
gauge invariance. The breaking of gauge invariance is removed by taking the 
masses of the Pauli-Villars electrons to infinity after the calculation is 
complete. It is not obvious that taking such a limit really restores gauge 
invariance but it can be shown to be true\cite{rf:10b}. 

In the present calculation we keep one 
Pauli-Villars photon and investigate how many Pauli-Villars 
electrons are required to successfully regularize the one loop electron 
self energy. We find that we need two Pauli-Villars electrons and so we 
use the Lagrangian
\begin{equation}
 \sum_{i=0}^1 (-1)^{i+1}{1 \over 4}{F^i}^{\mu \nu} {F^i}_{\mu \nu}
  - B A_- + \sum_{i} {1 \over \nu_i}\bar{\psi_i} (i \gamma^\mu \partial_\mu
  - m_i) \psi_i - e \bar{\psi}\gamma^\mu \psi A_\mu,
\end{equation}
where
\begin{equation}
   \psi = \sum_{i=0}^2 \psi_i , \quad \sum_{i=0}^2 \nu_i  = 0 ,
      \quad A_\mu = \sum_{i=0}^1 A^i_\mu.
\end{equation}
We shall have $m_0$ equal to the physical electron mass and $\nu_0 = 1$.  For 
the other PV condition we shall require that $\{\nu_i\}$ and $\{m_i\}$ be 
chosen such that the requirement from chiral symmetry --- that the renormalized mass be zero if the bare mass is zero --- be satisfied.  That is, we shall 
require that 
\begin{equation}
   \delta m|_{m_0 = 0} = 0 .
\end{equation}
Chiral transformations are dynamic in light-cone quantization and it is common 
that that requirement from chiral symmetry has to be imposed with an extra 
Pauli-Villars field\cite{rf:16}. As is usual, we shall regulate the infrared 
singularities with photon masses and shall add these to regulate the 
propagator rather than add a mass term to regulate the Lagrangian. 

Using  eqn's (\ref{eq:5.34}), (\ref{eq:5.36}), (\ref{eq:5.40}) 
and (\ref{fp}) we find that the one loop electron self energy is given by
\begin{equation}
   \frac{e^2}{(2 \pi)^4} \int d^4q \sum_{i,j}(-1)^{j} \nu_i D_{\mu \nu}
(q,\mu_j) \bra{p} \gamma^0 
\gamma^\mu \{\frac{(p-q)_\rho \gamma^\rho +m_i}{(p-q)^2 - m_i^2 +i\varepsilon} 
- \ha \frac{\gamma^+}{(p-q)^+} \}\gamma^\nu \ket{p}  .
\end{equation}
We shall take $p_\perp = 0$.  Due to the fact that they do not depend on 
masses, the contributions from the noncovariant term in the fermion propagator 
(the second term in (\ref{fp})), the contributions from the contact term in the
 gauge propagator (the last term in (\ref{eq:5.40})) and the contributions 
from the second and third terms in (\ref{eq:5.45}) all go to zero in the sum 
over $i$ and $j$. We can write the gauge propagator as
\begin{equation}
D_{\mu\nu}(q,\mu)=i\frac{-g_{\mu\nu}+\frac{n_{\mu}q_{\nu}+n_{\nu}q_{\mu}}
{q_-+i\varepsilon {\rm sgn}(q_+)}}{q^2-\mu^2+i\varepsilon}
\end{equation}
We then get
\begin{equation}
    {\Sigma}(p)=\frac{ie^2}{(2{\pi})^4}\!\!\int\!\!d^4q \sum_{i,j}\nu_i (-1)^j 
\; \frac{\gamma^{\mu}\{{\gamma}\cdot(p-q)+m_i\}\gamma^{\nu}}
{(p-q)^2-m_i^2+i{\varepsilon}}{\cdot}
\frac{-g_{\mu\nu}+\frac{n_{\mu}q_{\nu}+n_{\nu}q_{\mu}}{q_-+i{\varepsilon}
{\rm sgn}(q_+)}}{q^2-{\mu}_j^2+i{\varepsilon}}.
\end{equation}
We divide this expression into two pieces, $\Sigma^{(1)}(p)$ and 
$\Sigma^{(2)}(p)$, according to
\begin{eqnarray}
{\Sigma}^{(1)}(p)&=&\frac{-ie^2}{(2{\pi})^4}\int d^4q \sum_{i,j}\nu_i (-1)^j \;
 \frac{\gamma^{\mu}\{{\gamma}\cdot(p-q)+m_i\}\gamma_{\mu}}
{[(p-q)^2-m_i^2+i{\varepsilon}][q^2-{\mu}_j^2+i{\varepsilon}]} \\
{\Sigma}^{(2)}(p)&=&\frac{ie^2}{(2{\pi})^4}\int d^4q \sum_{i,j}\nu_i (-1)^j \;
 \frac{\gamma^{\mu}\{{\gamma}\cdot(p-q)+m_i\}\gamma^{\nu}}
{(p-q)^2-m_i^2+i{\varepsilon}}
\frac{\frac{n_{\mu}q_{\nu}+n_{\nu}q_{\mu}}{q_-+i{\varepsilon}
{\rm sgn}(q_+)}}{q^2-{\mu}_j^2+i{\varepsilon}}.
\end{eqnarray}
We now perform the $\gamma$ algebra and take the inner product  
$\bra{p}\gamma^0\Sigma^{(1)}(p)\ket{p}$ to get 
\begin{equation}
   \delta m^{(1)} = \frac{i e^2}{(2 \pi)^4} \int d^4q \sum_{i,j}\nu_i (-1)^j 
\frac{2 \frac{p^+}{m_0}(p_+ - q_+) + 2 \frac{p^-}{m_0}(p_- - q_-) - 4 m_i}
{[(p - q)^2 - m_i^2 + i\varepsilon][q^2 - \mu_j^2 + i \varepsilon]} .
\end{equation}
Here it is important to note that when we perform the $q_+$ integration, we 
obtain an extra contribution from the semicircle at infinity to the term 
propotional to $(p_+-q_+)$ as in the following
\begin{eqnarray*}
&&\int_{-\infty}^{\infty}dq_+\frac{(p-q)_+}{[(p-q)^2-m_i^2+i\varepsilon]
[q^2-\mu_j^2+i\varepsilon]} \nonumber \\
&=&\frac{-i\pi}{4q_-(p-q)_-}\left[\frac{\{{\rm sgn}(q_-)+{\rm sgn}((p-q)_-)\}
\frac{(p-q)_{\perp}^2+m_i^2}{2(p-q)_-}}{p_+-\frac{(p-q)_{\perp}^2+m_i^2-
i\varepsilon}{2(p-q)_-}-\frac{q_{\perp}^2+\mu_j^2-i\varepsilon}{2q_-}}
+{\rm sgn}(q_-)\right].
\end{eqnarray*}
and that it gives rise to a logarithmic divergence
$$\int_{-\infty}^{\infty}dq_-\frac{{\rm sgn}(q_-)}{4q_-(p-q)_-}
=\frac{1}{2p_-}\int_0^1dx \frac{1}{1-x}$$
so that a similar divergence resulting from $\frac{(p-q)_{\perp}^2+m_i^2}
{2(p-q)_-}$ is cancelled. As a consequence, changing the integration variable 
from $q_-$ to $x=\frac{q_-}{p_-}$ yields 
\begin{eqnarray*}
&&\delta m^{(1)}=\frac{e^2}{(2\pi)^3}\int d^2q_{\perp}\int_0^1dx\sum_{i,j}
\nu_i(-1)^j\frac{m_0(1-x)-2m_i}{m_0^2x(1-x)-m_i^2x-\mu_j^2(1-x)-q_{\perp}^2}  
\nonumber \\
&+&\frac{e^2}{2(2\pi)^3m_0}\int d^2q_{\perp}\int_0^1dx\sum_{i,j}
\nu_i(-1)^j\frac{m_0(2x-1)+m_i^2-\mu_j^2}
{m_0^2x(1-x)-m_i^2x-\mu_j^2(1-x)-q_{\perp}^2} .
\end{eqnarray*}
By performing the $x$ integration of the term in the second line we obtain 
\begin{eqnarray*}
&{}&\int d^2q_{\perp}\int_0^1dx\sum_{i,j}
\nu_i(-1)^j\frac{m_0(2x-1)+m_i^2-\mu_j^2}
{m_0^2x(1-x)-m_i^2x-\mu_j^2(1-x)-q_{\perp}^2}  \nonumber \\
&=&-\int d^2q_{\perp}\sum_{i,j}
\nu_i(-1)^j\log\left(\frac{q_{\perp}^2+m_i^2}{q_{\perp}^2+\mu_j^2}\right)
=-\int d^2q_{\perp}\sum_{i}
\nu_i\log\left(\frac{q_{\perp}^2+\mu_1^2}{q_{\perp}^2+\mu_0^2}\right).
\end{eqnarray*}
We see here that, due to the Pauli-Villars photon, this term is 
independent of Pauli-Villars electron masses so that if we impose the 
Pauli-Villars condition $\sum_{i}\nu_i=0$, it is trivially zero.
Consequently we obtain
\begin{equation}
\delta m^{(1)}=\frac{e^2}{(2\pi)^3}\int d^2q_{\perp}\int_0^1dx\sum_{i,j}
\nu_i(-1)^j\frac{m_0(1-x)-2m_i}{m_0^2x(1-x)-m_i^2x-\mu_j^2(1-x)-q_{\perp}^2}.  
\end{equation} 
For terms involving only the physical electron mass we have
\begin{equation}
\delta m^{(1)}_{physical} = \frac{ e^2}{(2 \pi)^3} \int d^2q_\perp \int_0^1 dx 
\sum_{j} (-1)^j  \frac{ m_0(1+x)}{ m_0^2 x^2 + \mu_j^2 (1-x) + q_\perp^2 }.
\end{equation}
The remaining contributions from $\Sigma^{(1)}$ are given by
\begin{equation}
 \delta m^{(1)}_{PV} = \frac{ e^2}{(2 \pi)^3 } \int d^2q_\perp \int_0^1 dx 
\sum_{i=1}^2 \sum_{j} \nu_i(-1)^j  \frac{ m_0 (1-x) - 2 m_i }
{ m_0^2 x (1-x) - m_i^2 x - \mu_j^2 (1-x) - q_\perp^2 }.
\end{equation}
It is not difficult to show that this quantity goes to zero as the masses of 
the Pauli-Villars electrons go to infinity. In this way we obtain the previous result 
that one Pauli-Villars electron and one Pauli-Villars photon are enough to 
regularize $\delta m^{(1)}$.

We now consider the contributions from $\Sigma^{(2)}$.  Taking the matrix 
element, we find that
\begin{eqnarray}
\delta m^{(2)}&=&-\frac{ie^2}{(2{\pi})^4}\frac{2p^+}{m_0}\int d^4q \sum_{i, j} 
\nu_i (-1)^{j}\frac{1}{[q^2-{\mu}^2_j+i{\varepsilon}][q_-+i{\varepsilon}
{\rm sgn}(q_+)]} \nonumber \\
&-&\frac{ie^2}{(2{\pi})^4}\int d^4q \sum_{i=1}^2 \sum_{ j} \nu_i(-1)^{j} 
\left(\frac{2(m_0-m_i)}{[(p-q)^2-m_i^2+i{\varepsilon}][q^2-{\mu}^2_j
+i{\varepsilon}]} \right. \nonumber \\
&-& \left.\frac{2\frac{m_0^2-m_i^2}{m_0}p^+}
{[(p-q)^2-m_i^2+i{\varepsilon}][q^2-{\mu}^2_j+i{\varepsilon}]
[q_-+i{\varepsilon}{\rm sgn}(q_+)]} \right) .
\end{eqnarray}
The first term is trivially zero and the second term goes to zero as the 
masses of the Pauli-Villars electrons go to infinity. Thus we can discard them
 and concentrate on the third term, which is given by
\begin{equation}
\delta m^{(2)}=\delta m^{(2)}_1+\delta m^{(2)}_2
\end{equation}
where
\begin{eqnarray}
\delta m^{(2)}_1&=&\frac{-ie^2}{(2{\pi})^4}\int d^2q_{\perp}\int\frac{dq_-}
{4q_-(p-q)_-} \sum_{i,j} \frac{2\nu_i(-1)^{j}\frac{m_0^2-m_i^2}{m_0}p^+}
{[p_+--\frac{(p-q)_{\bot}^2+m^2-i{\varepsilon}}{2(p-q)_-}
-\frac{q_{\bot}^2+{\mu}^2-i{\varepsilon}}{2q_-}]} \nonumber \\
&\times&\frac{q_-}{q_-^2+\varepsilon^2}\int_{-\infty}^{\infty}
dq_+\left(\frac{1}{q_+-p_++\frac{(p-q)_{\bot}^2
+m^2-i{\varepsilon}}{2(p-q)_-}}-\frac{1}{q_+-\frac{q_{\bot}^2+\mu_j^2-
i{\varepsilon}}{2q_-}}\right) \\
\delta m^{(2)}_2&=&\frac{ie^2}{(2{\pi})^4}\int d^2q_{\perp}\int\frac{dq_-}
{4q_-(p-q)_-} \sum_{i,j} \frac{2\nu_i(-1)^{j}\frac{m_0^2-m_i^2}{m_0}p^+}
{[p_+--\frac{(p-q)_{\bot}^2+m^2-i{\varepsilon}}{2(p-q)_-}
-\frac{q_{\bot}^2+{\mu}^2-i{\varepsilon}}{2q_-}]} \nonumber \\
&\times&\frac{i\varepsilon}{q_-^2+\varepsilon^2}\int_{-\infty}^{\infty}
dq_+\left(\frac{1}{q_+-p_++\frac{(p-q)_{\bot}^2
+m^2-i{\varepsilon}}{2(p-q)_-}}-\frac{1}{q_+-\frac{q_{\bot}^2+\mu_j^2-
i{\varepsilon}}{2q_-}}\right).
\end{eqnarray}
By performing the $q_+$ integration $\delta m^{(2)}_1$ and 
$\delta m^{(2)}_2$ can be described, respectively, as  
\begin{eqnarray}
\delta m^{(2)}_1&=&\frac{-e^2}{(2{\pi})^3}\int d^2q_{\perp}\int_0^1dx
\sum_{i,j} \frac{\nu_i(-1)^{j}\frac{m_0^2-m_i^2}
{m_0}\frac{x}{x^2+\varepsilon^2}}
{q_{\perp}^2+m_i^2x+\mu_j^2(1-x)-m_0^2x(1-x)} \\
\delta m^{(2)}_2&=& \frac{e^2}{(2{\pi})^4}\int d^2q_{\perp}\int dq_- 
\frac{\varepsilon}{q_-^2+\varepsilon^2}\sum_{i,j} \nonumber \\ 
&{\times}&\frac{\nu_i(-1)^{j}\frac{m_0^2-m_i^2}{m_0}p^+ (L+2iT)}
{2p_+q_-(p_--q_-)-q_-(q_{\perp}^2+m_i^2-i \varepsilon)-(p_--q_-)
(q_{\perp}^2+\mu_j^2-i\varepsilon)}  
\end{eqnarray}
where
\begin{eqnarray}
L&=&\log\left(\frac{q_-^2(q_{\perp}^2+m_i^2-2p_+(p_--q_-))^2}
{(p_--q_-)^2(q_{\perp}^2+\mu_j2)^2}\right),\\
T&=&\tan^{-1}\left(\frac{\varepsilon(m_i^2-\mu_j^2-2p_+(p_--q_-))}
{(q_{\perp}^2+\mu_j^2)(q_{\perp}^2+m_i^2-2p_+(p_--q_-))}\right).
\end{eqnarray}
Here we see that the term proportional to T is well-defined so that the factor 
$\frac{\varepsilon}{q_-^2+\varepsilon^2}$ behaves as 
$\pi\delta(q_-)$. As a result, it vanishes as $\varepsilon$ tends to 
zero. Therefore we can discard it. The remaining integral violates the chiral symmetry condition.  That is why we need the second Pauli-Villars electron.  We impose the last 
Pauli-Villars condition --- $ \delta m|_{m_0 = 0} = 0$. 
Actually, we see that $\delta m^{(2)}$ diverges as $m_0 \rightarrow 0$; so we 
set 
\begin{eqnarray}
&&\frac{ e^2}{(2{\pi})^3}\int d^2q_{\perp}\sum_{i,j}\nu_i(-1)^{j}
\left[\int_0^1dx \frac{x}{x^2+\varepsilon^2}\frac{\frac{m_0^2-m_i^2}
{m_0}}{q_{\perp}^2+m_i^2x+\mu_j^2(1-x)} \right. \nonumber \\
&+& \left.\frac{1}{\pi}\int dq_- \frac{\varepsilon}{q_-^2+\varepsilon^2} 
\frac{\frac{m_0^2-m_i^2}{m_0}p^+ 
\log\left|\frac{q_-(q_{\perp}^2+m_i^2)}
{(p_--q_-)(q_{\perp}^2+\mu_j^2)}\right|}
{q_-(q_{\perp}^2+m_i^2-i \epsilon)+(p-q)_-(q_{\perp}^2+\mu_j^2-i\epsilon)} 
\right]=0.
\end{eqnarray}
We can use this relation to write $\delta m^{(2)}$ as 
\begin{equation}
\delta m^{(2)}=\delta m^{(2)}_{c1}+\delta m^{(2)}_{c2}
\end{equation}
where 
\begin{eqnarray}
&&\delta m^{(2)}_{c1}=-\frac{e^2}{(2{\pi})^3}\int d^2q_{\perp}
\sum_{i,j}\nu_i (-1)^{j}\int_0^1dx \frac{x}{x^2+\varepsilon^2}
\frac{m_0^2-m_i^2}{m_0}\nonumber \\
&{\times}&\left(
\frac{1}{q_{\perp}^2+m_i^2x+\mu_j^2(1-x)-m_0^2x(1-x)}
-\frac{1}{q_{\perp}^2+m_i^2x+\mu_j^2(1-x)}\right) \nonumber \\
&=&-\frac{e^2}{8{\pi}^2}
\sum_{i,j}\nu_i (-1)^{j}\!\!\int_0^1\!\!dx \frac{x}{x^2+\varepsilon^2}
\frac{m_i^2-m_0^2}{m_0}\log\left(1-\frac{m_0^2x(1-x)}{m_i^2x+\mu_j^2(1-x)}
\right)\!\!, \label{eq:5.69}
\end{eqnarray}
\begin{eqnarray}
&&\delta m^{(2)}_{c2}= \frac{e^2}{(2{\pi})^3}\int d^2q_{\perp} \frac{1}{\pi}
\int dq_- \frac{\varepsilon}{q_-^2+\varepsilon^2}\sum_{i,j}\nu_i(-1)^{j} 
\nonumber\\
&{\times}&\left[\frac{\frac{m_0^2-m_i^2}{m_0}p^+ 
\log\left(1-\frac{2p_+(p_--q_-)}
{q_{\perp}^2+m_i^2}\right)}
{2p_+q_-(p_--q_-)-q_-(q_{\perp}^2+m_i^2-i \varepsilon)-(p_--q_-)
(q_{\perp}^2+\mu_j^2-i\varepsilon)}\right. \nonumber \\  
&+&\left.2m_0(m_0^2-m_i^2)q_-(p_--q_-) 
\log\left|\frac{q_-(q_{\perp}^2+m_i^2)}{(p_--q_-)(q_{\perp}^2+\mu_j^2)}\right|
\right. \nonumber \\
&\times&\left.\frac{\{q_-(q_{\perp}^2+m_i^2)+(p_--q_-)(q_{\perp}^2+\mu_j^2)\}^
{-1}}{2p_+q_-(p_--q_-)-q_-(q_{\perp}^2+m_i^2)-(p_--q_-)(q_{\perp}^2+\mu_j^2)}
\right].
\end{eqnarray}
Now the integrand of $\delta m^{(2)}_{c2}$ can be regarded as a continuous 
function of $q_-$. Consequently the factor $\frac{1}{\pi}\frac{\varepsilon}
{q_-^2+\varepsilon2}$ behaves  as $\delta(q_-)$ so that we obtain
\begin{eqnarray}
\lim_{\varepsilon{\to}0}\delta m^{(2)}_{c2}
&=&-\frac{e^2}{(2{\pi})^3}\int d^2q_{\perp} \sum_{i,j}\nu_i(-1)^{j}
\frac{\frac{m_0^2-m_i^2}{m_0} \log\left(1-\frac{m_0^2}
{q_{\perp}^2+m_i^2}\right)}{q_{\perp}^2+\mu_j^2}. \label{eq:5.71}
\end{eqnarray}
It follows from (\ref{eq:5.69}) and (\ref{eq:5.71})) that
\begin{equation}
\lim_{m_i\to\infty}\left(\lim_{\varepsilon{\to}0}\delta m^{(2)}\right)=0.
\end{equation}

  So the final answer for the electron self energy is regulated by one 
Pauli-Villars photon and is given by
\begin{equation}
\delta m = \frac{ e^2}{(2 \pi)^3 m_0} \int d^2q_\perp \int_0^1 dx 
\sum_{j} (-1)^j \frac{ m_0^2 (1+x)}{ m_0^2 x^2 + \mu_j^2 (1-x) + q_\perp^2 }.
\end{equation}
That is the same answer as the one obtained by Feynman methods
\cite{rf:10b,rf:18}.

We end this section by making some remarks. If we regulate the spurious 
singularity instead by 
\begin{equation}
\frac{1}{q_-+iq_+\varepsilon}, 
\end{equation}
which has been often used as the ML prescription, and if we take $\varepsilon$ 
to zero at the end of the calculation, then it improves properties of the 
$q_+$ integration so that we need not introduce extra Pauli-Villars electrons. 
It is also possible to include a regulator in the spurious pole of the form 
$\frac{1}{q_- + \epsilon}$; in that case $\varepsilon$ can be taken to zero 
immediately after the $q_+$ integration is performed and, if we keep 
$\epsilon$ finite until after the Pauli-Villars electron masses are taken to 
infinity, we again obtain the Feynman answer. The main point is that the 
cancellation of the strongest singularity in the gauge propagator, shown in 
(\ref{eq:5.39}), allows a successful calculation of the electron self energy 
using standard techniques.

\section{Concluding remarks}
In this paper we have constructed the quantization of QED in gauges fixed 
by specifying a constant, space-like 
vector, $n$, and the gauge condition, $n^{\mu}A_{\mu}=0$. We have then 
constructed 
the axial gauge formulation of QED in the $\pm$-coordinates. 
Our framework has allowed us to consider axial gauges generally. The 
temporal and axial gauges in ordinary coordinates correspond, respectively, to 
$\theta=0$ and $\theta=\frac{\pi}{2}$, while the light-cone formulation 
corresponds to $\theta=\frac{\pi}{4}$. The most important aspect of this 
framework is that it has enabled us to use 
the temporal gauge formulation to obtain the algebra of the RG fields, 
which are not canonical variables in the pure space-like case; they 
are nevertheless, necessary ingredients in that case.

We have obtained the commutation relations of $B$ in the temporal 
region and extrapolated them into the axial region. We have also obtained the 
formal solution (\ref{eq:3.8}) of the gauge field equations and specified the 
relevant integration constants by comparing with the free gauge fields 
in the temporal formulation. We have made use of that solution to specify the 
integration constants that appear when we solve the constraint equations 
(\ref{eq:3.12}) and (\ref{eq:3.13}) in favor of the independent canonical 
variables. We have specified the equal $x^+$-time commutation relations of the 
constituent fields as being canonical and found eventually that the RG 
fields make vanishing 
contributions to the equal $x^+$-time commutation relations because 
they are multiplied by the operator $\frac{1}{{\nabla}_T^{\;\;2}}$. We have 
furthermore obtained the conserved Hamiltonian to which Hamiltonian for the RG 
fields is added. In this way we have succeeded in constructing a perturbative 
formulation of pure space-like axial gauge QED in which the RG fields 
regularize infrared divergences inherent in the traditional quantizations of 
those gauges. The resulting gauge field propagator has the ML form. 

We have illustrated the effect of the RG fields by performing a calculation of 
the one loop self energy of the electron.  In the usual light-cone gauge 
without higher derivatives so that $A_+$ is a constrained field, that 
calculation has not previously been performed successfully.  The severe 
infrared divergences which result from solving the constraint equation for 
$A_+$ (without inclusion of the integration constants which we label $B$ and 
$C$) have prevented a successful calculation.  What we have found here is that 
including the integration constants (the RG fields) softens these severe 
infrared singularities and allows a successful calculation of the self energy 
using standard regulation techniques.  The necessary regulation is slightly 
more complicated than in Feynman gauge but is considerably simpler than in the 
higher derivative regulated version of light-cone gauge where $A_+$ is a degree
 of freedom.  Indeed we expect that the very complicated regulation procedures 
 necessary in light-cone gauge without the RG fields \cite{rf:10b} are somehow 
mimicking the effects of the RG fields, which should be included, to a 
sufficient degree that an effective renormalization can be performed. We should
 also remark that if another order of performing the integral is used (such as 
performing the $q_-$ integral first) then a smaller number of regulator fields 
is needed.  We have studied the case of performing the $q_+$ integral first 
since that is the order that makes closest contact with nonperturbative 
light-cone calculations.

 While we have only considered the case of one loop (with the $q^+$ integral 
performed first), we expect that it may be possible to use the techniques of 
Paston and Franke\cite{rf:16} to show that the calculations with the relatively
 simple regulation are equivalent to Feynman methods to all orders.  Since the 
quantization of QCD in Feynman gauge encounters difficulties which have not, so
 far, been solved, it may be that if the RG fields can be included in QCD, as 
they were here, the resulting formulation would have practical advantages over 
other formulations.  


\appendix
\section{ Verification of {\rm ~Eq.~(\ref{eq:5.31})}}

We change the integration variable from $q_-$ to 
$k_+=\frac{n_+q_--\sqrt{q_-^2-n_-q_{\bot}^2}}{n_-}$. The quantity $k_+$ is a 
two-valued function of $q_-$, and it takes its minimum value, $\sqrt{-n_-}
q_{\bot}{\equiv}m_0$, at $q_-=\frac{n_+}{-n_-}m_0$. Therefore, when we change 
the integration variable from $q_-$ to $k_+$, we use $q_-=
\frac{n_+k_+-\sqrt{k_+^{\;2}-m_0^{\;2}}}{-n_-}$ so that the region 
$m_0{\leqq}k_+<\infty$ corresponds to $-\infty<q_-{\leqq}\frac{n_+}{-n_-}
m_0$, whereas  we use $q_-=\frac{n_+k_++\sqrt{k_+^{\;2}-m_0^{\;2}}}{-n_-}$ so 
that the region $m_0{\leqq}k_+<\infty$ corresponds to $\frac{n_+}
{-n_-}m_0{\leqq}q_-<\infty$. Hence we have 
\begin{eqnarray} 
&{}&\int_{-\infty}^{\infty}\frac{dq_-}{q_-^{\;2}}
=\int_{m_0}^{\infty}dk_+
\frac{k_+-n_+\sqrt{k_+^2-m_0^2}}{(-n_-)\sqrt{k_+^{\;2}-m_0^{\;2}}}
\left(\frac{-n_-}{n_+k_+-\sqrt{k_+^{\;2}-m_0^{\;2}}}\right)^2 \nonumber \\
&+&\int_{m_0}^{\infty}dk_+
\frac{k_++n_+\sqrt{k_+^2-m_0^2}}{(-n_-)\sqrt{k_+^{\;2}-m_0^{\;2}}}
\left(\frac{-n_-}{n_+k_++\sqrt{k_+^{\;2}-m_0^{\;2}}}\right)^2,
\label{eq:A.1}
\end{eqnarray} 
where the first term diverges, but the second term is finite. Then, combining 
the first integral with the second one of (\ref{eq:5.31}) and making use of 
an equality
\begin{eqnarray}
&{}&(k_+-n_+\sqrt{k_+^2-m_0^2})(n_+k_++\sqrt{k_+^{\;2}-m_0^{\;2}})^2
-2n_+n_-\sqrt{k_+^{\;2}-m_0^{\;2}}(n_-k_+^2-q_{\bot}^2) \nonumber \\
&=&(k_++n_+\sqrt{k_+^2-m_0^2})(n_+k_+-\sqrt{k_+^{\;2}-m_0^{\;2}})^2
\end{eqnarray}
yields
\begin{eqnarray}
&{}&\int_{m_0}^{\infty}\!\!dk_+
\frac{k_+-n_+\sqrt{k_+^2-m_0^2}}{(-n_-)\sqrt{k_+^{\;2}-m_0^{\;2}}}
\!\left(\frac{-n_-}{n_+k_+-\sqrt{k_+^{\;2}-m_0^{\;2}}}\right)^2\!
+\!{\int}^{\infty}_0\!\!dk_+\frac{2n_+(n_-k_+^2-q_{\bot}^2)}
{(n_-k_+^2+q_{\bot}^2)^2} \nonumber \\
&=&\int_{m_0}^{\infty}\!\!dk_+
\frac{k_+-n_+\sqrt{k_+^2-m_0^2}}{(-n_-)\sqrt{k_+^{\;2}-m_0^{\;2}}}
\!\left(\frac{n_+k_++\sqrt{k_+^{\;2}-m_0^{\;2}}}{n_-k_+^{\;2}+q_{\bot}^2}
\right)^2\!
+\!\int_0^{\infty}\!\!dk_+\frac{2n_+(n_-k_+^2-q_{\bot}^2)}{(n_-k_+^{\;2}
+q_{\bot}^2)^2} 
\nonumber \\
&=&\!\int_{m_0}^{\infty}\!\!dk_+
\frac{k_++n_+\sqrt{k_+^2-m_0^2}}{(-n_-)\sqrt{k_+^{\;2}-m_0^{\;2}}}
\!\left(\frac{n_+k_+-\sqrt{k_+^{\;2}-m_0^{\;2}}}
{n_-k_+^{\;2}+q_{\bot}^2}\right)^2\!
+\!\int_0^{m_0}\!\!dk_+\frac{2n_+(n_-k_+^2-q_{\bot}^2)}
{(n_-k_+^{\;2}+q_{\bot}^2)^2}. \label{eq:A.3}\nonumber \\
\end{eqnarray}
We see that the first integral in the last line is identical with the second 
one in (\ref{eq:A.1}) and thus is finite. Carrying out the integrations, we 
obtain
\begin{eqnarray}
&2\int_{m_0}^{\infty}dk_+
\frac{k_++n_+\sqrt{k_+^2-m_0^2}}{(-n_-)\sqrt{k_+^{\;2}-m_0^{\;2}}}
\left(\frac{-n_-}{n_+k_++\sqrt{k_+^{\;2}-m_0^{\;2}}}\right)^2
=\frac{\sqrt{-n_-}}{n_+}\frac{2}{q_{\bot}},\label{eq:A.4}& \\
&\int_0^{m_0}dk_+\frac{2n_+(n_-k_+^2-q_{\bot}^2)}
{(n_-k_+^{\;2}+q_{\bot}^2)^2}=-\frac{\sqrt{-n_-}}{n_+}\frac{2}{q_{\bot}}.&
\label{eq:A.5}
\end{eqnarray}
It follows that 
\begin{equation}
{\int}^{\infty}_{\!\!\!-\infty}dq_-\frac{1}{q_-^2}+{\int}_0^{\infty}dk_+
\frac{2n_+(n_-k_+^2-q_{\bot}^2)}
{(n_-k_+^2+q_{\bot}^2)^2}={\rm (\ref{eq:A.4})}+{\rm (\ref{eq:A.5})}=0.  
\end{equation}



\begin{thebibliography}{99}
\bibitem{rf:1}
G.~Leibbrandt, Rev. Mod. Phys. \andvol{59,1987, 1067}.\\
A.~Bassetto,~G.~Nardelli and R.~Soldati, {\it Yang-Mills Theories in Algebraic
 Non-Covariant Gauges} (World Scientific, Singapore, 1991).
\bibitem{rf:2}
Yu.~L.~Dokshitzer,~D.~J.~D'Yakonov, and S.~Y.~Troyan, Phys. Rep.
\andvol{58,1980,270}.
\bibitem{rf:3}
S.~J.~Brodsky,~H.~C.~Pauli and S.~S.~Pinsky, Phys. Rep. \andvol{301,1998, 299}.
\bibitem{rf:4}
N.~Nakanishi, 
        \JL{Phys.~Lett.,131B,1983, 381}.   
\\
N.~Nakanishi, {\it Quantum Electrodynamics},~ed.T.~Kinoshita 
(World Scientific, Singapore, 1990), p.~36.
\bibitem{rf:5}
D.~M.~Capper, J.~J.~Dulwich, and M.~J.~Litvak, \NP{241B,1984,463}. \\
S.~Mandelstam, 
        Nucl.~Phys.\ {\bf B213} (1983), 149. \\
G.~Leibbrandt, Phys.~Rev.\ {\bf D29} (1984), 1699.   
\bibitem{rf:6}
A.~Bassetto, ~M.~Dalbosco,~I.~Lazziera and R.~Soldati, 
        Phys.~Rev.\ {\bf D31} (1985), 2012.   

\bibitem{rf:7}
I.~Lazzizzera, \PL{210B,1988,188}; Nuovo Cim. \andvol{102A,1989,1385}. \\
M.~Morara and R.~Soldati, \PR{D58,1998,105011}.   
\bibitem{rf:8}
G.~McCartor and D.G.~Robertson, 
Z.~Phys.~\andvol{C62,1994, 349}; \andvol{C68,1995,345}. 
\bibitem{rf:9}
Y.~Nakawaki and G.~McCartor, \PTP{106,2001,167}. \\
Y.~Nakawaki and G.~McCartor, \PTP{111,2004,}.
\bibitem{rf:10}
K.~Hornbostel, \PR{D45,1992,3781}; E.~Prokhvatilov and V.~Franke, Yad. Fiz. \andvol{49,1989,1109}; F.~Lenz, M.~Thies, S.~Levit, and D.~Yazaki, Ann. Phys. \andvol{208,1991,1}.
\bibitem{rf:10a}  
A. Langnau and M. Burkardt, \PR{D47,1993,3452}.
\bibitem{rf:10b}
S. J. Brodsky, V. A. Franke, J. R. Hiller, G. McCartor, S. A. Paston and
E. V. Prokhvatilov,~Nucl.~Phys.\ {\bf B703} (2004), 333.
\bibitem{rf:11}
G.~McCartor, Z.~Phys.~\andvol{C41,1988, 271}. 
\bibitem{rf:12}
B.~Lautrup, Mat.~Fys.~Medd.~Dan.~Vid.~Selsk.~{\bf 35}(1967),~No.11.  \\
N.~Nakanishi, Prog. Theor. Phys. Suppl. No.51~(1972),1.
\bibitem{rf:13}
Y.~Nakawaki and G.~McCartor, \PTP{102,1999,149}. 
\bibitem{rf:14}
P.~A.~M.~Dirac, {\it Lectures on Quantum Mechanics}, Belfer Graduate 
School of Science-Yeshiva University (Academic Press, New York, 1964).
\\
A.~Hanson,~T.~Regge and C.~Teitelboim, {\it Constrained Hamiltonian Systems} \\
(Accad.~Naz.~deiLincei, Rome, 1976).
\bibitem{rf:15}
J.~Schwinger, \PR{130,1963,402}. 
\bibitem{rf:16}
S.-J.~Chang and T.-M.~Yan, \PR{D7,1973,1147}; \\
S.J. Brodsky, J.R. Hiller, and G. McCartor,\PR{D58,1998,025005} \\
S.A.~Paston and V.A.~Franke, Theor.~Math.~Phys.~\andvol{112,1997,1117}  \\
S.A.~Paston, V.A.~Franke, and E.V.~Prokhvatilov, Theor.~Math.~Phys.~\andvol{120,1999,1164}
\bibitem{rf:17}
S.J. Brodsky, J.R. Hiller, and G. McCartor, Ann.~Phys.~\andvol{305,2003,266}.
\bibitem{rf:18}
S.J. Brodsky, R. Roskies, and R. Suaya, \PR{D8,1973,4574}.
\end{thebibliography}
\end{document}